\documentclass[12pt]{iopart}
\usepackage[utf8]{inputenc} 
\usepackage{graphicx} 
\usepackage{float}
\usepackage{cite}
\usepackage{amssymb}
\usepackage{hyperref}
\usepackage{mhchem}
\usepackage[numbers]{natbib}
\setcitestyle{square}



\begin{document}

\title[Measurements of CMFX Fusion Yield]{Measurements of Fusion Yield on the Centrifugal Mirror Fusion Experiment}

\author{J. L. Ball$^1$, 
S. Mackie$^1$,  J. G. van de Lindt$^1$, W. Morrissey$^2$, A. Perevalov$^2$,  Z. Short$^3$, N. Schwartz$^3$, T. Koeth$^3$, B. L. Beaudoin$^3$, C. A. Romero-Talamás$^2$, J. Rice$^1$, R. A. Tinguely$^1$}

\address{$^1$MIT Plasma Science and Fusion Center; Cambridge, MA 02139}
\address{$^2$University of Maryland, Baltimore County; Baltimore, MD 21250}
\address{$^3$University of Maryland, College Park; College Park, MD 20742}

\ead{jlball@mit.edu}
\vspace{10pt}
\begin{indented}
\item[]May 2025
\end{indented}

\begin{abstract}
The Centrifugal Mirror Fusion Experiment (CMFX) at the University of Maryland, College Park is a rotating mirror device that utilizes a central cathode to generate a radial electric field which induces a strongly sheared azimuthal $E\times B$ flow to improve plasma confinement and stability. The fusion yield of CMFX plasmas is assessed by diagnosis of neutron emission for the first time. The total neutron yield is measured with two xylene (EJ-301) and deuterated-xylene (EJ-301D) liquid scintillator detectors absolutely calibrated with an \textit{in silico} method. A larger xylene scintillator was cross-calibrated and used to measure the time dynamics of the fusion rate under various experimental conditions. A permanently installed $^3$He gas tube detector was independently calibrated with a Cf-252 neutron source to make total yield measurements and provide an independent validation of the scintillator calibration. An interpretive modeling framework was developed using the 0D code MCTrans++ (Schwartz \textit{et al} 2024 JPP) to infer undiagnosed plasma parameters such as density, temperature, and confinement time. A peak neutron emission rate of 8.4$\times 10^{6}$ $\pm$ 7.0$\times 10^{5}$ was measured (neglecting modeling uncertainties), with an inferred triple product of 1.9~$\times~10^{17}$ $\mathrm{m^{-3}}$ keV s from 0D modeling.

\end{abstract}

%
%
%
%
%

\section{Introduction}
Centrifugal Mirrors are a class of fusion device which augments the confinement and performance of standard axisymmetric magnetic mirrors by applying a strong radial electric field to induce a supersonic, strongly sheared  azimuthal $E\times B$ flow \cite{Ellis2005}. MHD equilibria of the Centrifugal Mirror configuration are characterized by a balance between the plasma pressure, centrifugal force and magnetic tension; the effective centrifugal potential provides axial confinement which helps reduce end losses. The rotational shear suppresses the flute mode, a common MHD instability in simple mirrors, and improves the microscopic transport properties as well. In addition, the viscous heating associated with the rotational shear is theorized to be sufficient to achieve reactor-relevant plasma conditions, eliminating the need for auxiliary heating systems \cite{Ellis2001}. The physical principles of the Centrifugal Mirror concept have been demonstrated experimentally in several machines, beginning with the IXION experiment in the 1960s \cite{Baker1961}. More recent experiments include the PSP experiments which first demonstrated the complete suppression of the flute mode instability \cite{Volosov2009} and the Maryland Centrifugal Experiment (MCX) which studied the transport properties of centrifugal mirrors in detail \cite{Romero-Talamas2010, TeodorescuConfinementOfPlasmaVerification,Uzun-Kaymak2008}.

\begin{figure} [h]
    \centering
    \includegraphics[width=0.75\linewidth]{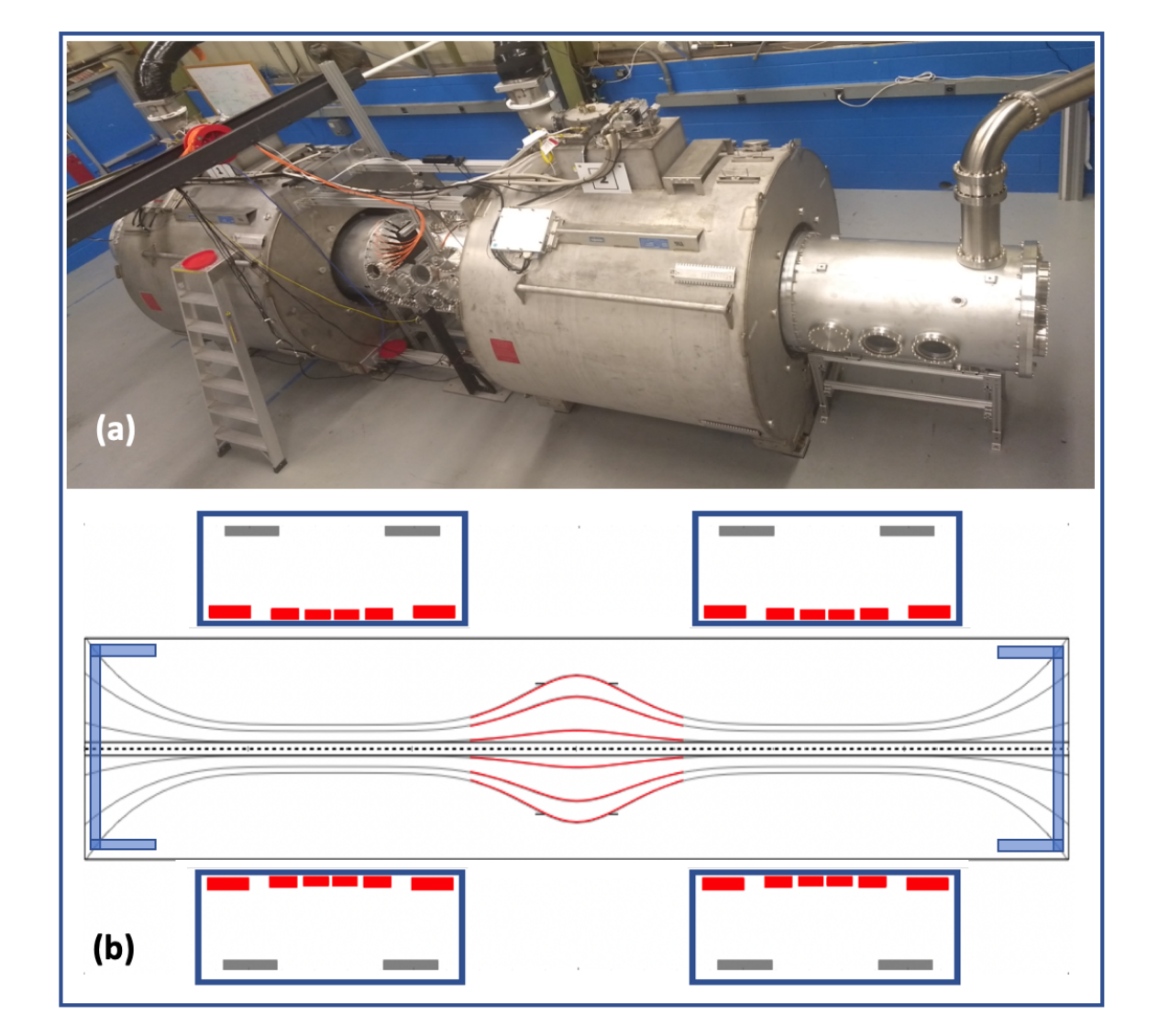}
    \caption{(a) CMFX superconducting magnets and vacuum vessel. (b) Schematic diagram of magnets with internal coils and resulting magnetic flux surfaces. Note that only flux surfaces where the plasma rotates are shown. The red portion of the contours correspond to the region of the volume measured with a tesla-meter, and were used to verify the calculated contours.}
    \label{fig:CMFXschematic}
\end{figure}

The Centrifugal Mirror Fusion Experiment (CMFX) is a new superconducting Centrifugal Mirror designed to investigate the scaling of such devices to reactor relevant conditions. First plasma was achieved in October 2022 using a capacitor bank. In May 2024, a DC power supply capable of delivering up to 100~kV and 100~kW replaced the capacitor bank. This latter power system was used for all the experiments reported here. The mirror field is 3~T at the throat and 0.3~T at midplane, giving a mirror ratio of 10. The magnetic field is produced by two large, low-temperature superconducting magnets separated by 1.45 m; these magnets are repurposed magnetic resonance imaging (MRI) magnets that were placed as close as possible to each other without compromising their mechanical integrity. The plasma vessel is 75~cm in diameter. A picture of the CMFX magnet and vacuum vessel, as well as a contour plot of the magnetic flux surfaces that confine the rotating plasma, are shown in Fig.~\ref{fig:CMFXschematic}. 

To avoid the rotating plasma scraping the vacuum vessel wall, the plasma is limited with tungsten-coated grounding electrodes with a diameter of 44~cm. While CMFX can operate with other gases, all plasmas studied in this work were fueled with pure deuterium. The power supply negatively biases an axial conductor relative to the machine vessel, providing average radial electric fields up to 454~kV/m capable of driving average $E\times B$ rotational flows up to 1500~km/s. CMFX exhibits steady and repeatable plasma conditions for bias voltages up to about 62.5 kV, above which arcing across the insulating standoff trips the power supply and prematurely terminates discharges. This arcing is found to be a consequence of the insulator design and not an intrinsic limitation of the centrifugal mirror concept. 

As a new fusion experiment, CMFX has relatively limited diagnostic capabilities. Several video cameras are mounted around the chamber and monitor the machine in the visible range with up to 700 fps. Two visible spectrometers are installed and monitor the spectra of the plasma in the central plane in the range between 350~nm and 1000~nm. CMFX is also equipped with a two-color interferometer with a CO2 laser ($\sim$ 10~$\mathrm{\mu}$m) and a diode pumped solid state laser (639~nm), but radiation heating of its ZnSe vacuum window interfered with the measurement during these experiments \cite{ZnSe_heating}. Deconvolution of the interferometer phase shift from window heating and plasma density is an active area of research and will be reported elsewhere. The high voltage DC power supply has a voltage and current monitor with 10~$\mathrm{\mu}$s time resolution, providing high quality data on the applied electric field, plasma current, and total input power. CMFX is equipped with permanently installed He-3 detectors to monitor the neutron yield. In this work, liquid scintillator detectors were temporarily installed to study the neutron emission in greater detail.

The remainder of this article is structured as follows: in Section \ref{sec:calib}, we describe our methodology for making absolutely calibrated measurements of CMFX neutron production; in Section \ref{sec:performance}, we report the measured fusion yield rate of CMFX under various experimental conditions; and in Section \ref{sec:modeling}, we develop an interpretive modeling framework to use measured yield, applied voltage, and input power to infer temperature, density, and confinement time. We conclude by summarizing and contextualizing the results.

\section{Calibrated Measurements of Fusion Neutron Production}
\label{sec:calib}
The neutron yield rate, $\dot{Y_n}$ with units of neutrons/s, is a critical figure of merit for assessing the performance of any fusion power reactor concept, as it is directly proportional to the total nuclear energy released by fusion reactions per second. In a thermonuclear plasma, a measurement of the neutron yield rate also constrains fuel ion density, $n_D$, and temperature, $T_D$, through the fusion reactivity, $\langle\sigma v\rangle$, which is a strong function of temperature \cite{Hively_fusion_cross_section_forms, Hutchinson_2002}.

Measurements of neutron rate, $\dot{M}$, are related to the total neutron yield rate by the deceptively simple equation:

\begin{equation}
\label{yield_eq}
    M=\epsilon \dot{Y_n}
\end{equation}

Here, $\epsilon$ is the detector's total efficiency and is defined as the ratio of measured neutrons to total neutrons produced by the plasma. Absolute calibration of yield measurements is the determination of this coupling factor, which is a complicated function of the experimental geometry, materials, neutron source profile and energy spectrum, and detector response function. In this work, we use two independent calibration methods to measure the fusion yield of CMFX plasmas and find good agreement between the methods, giving confidence in the results.

\begin{figure} [H]
    \centering
    \includegraphics[width=0.75\linewidth]{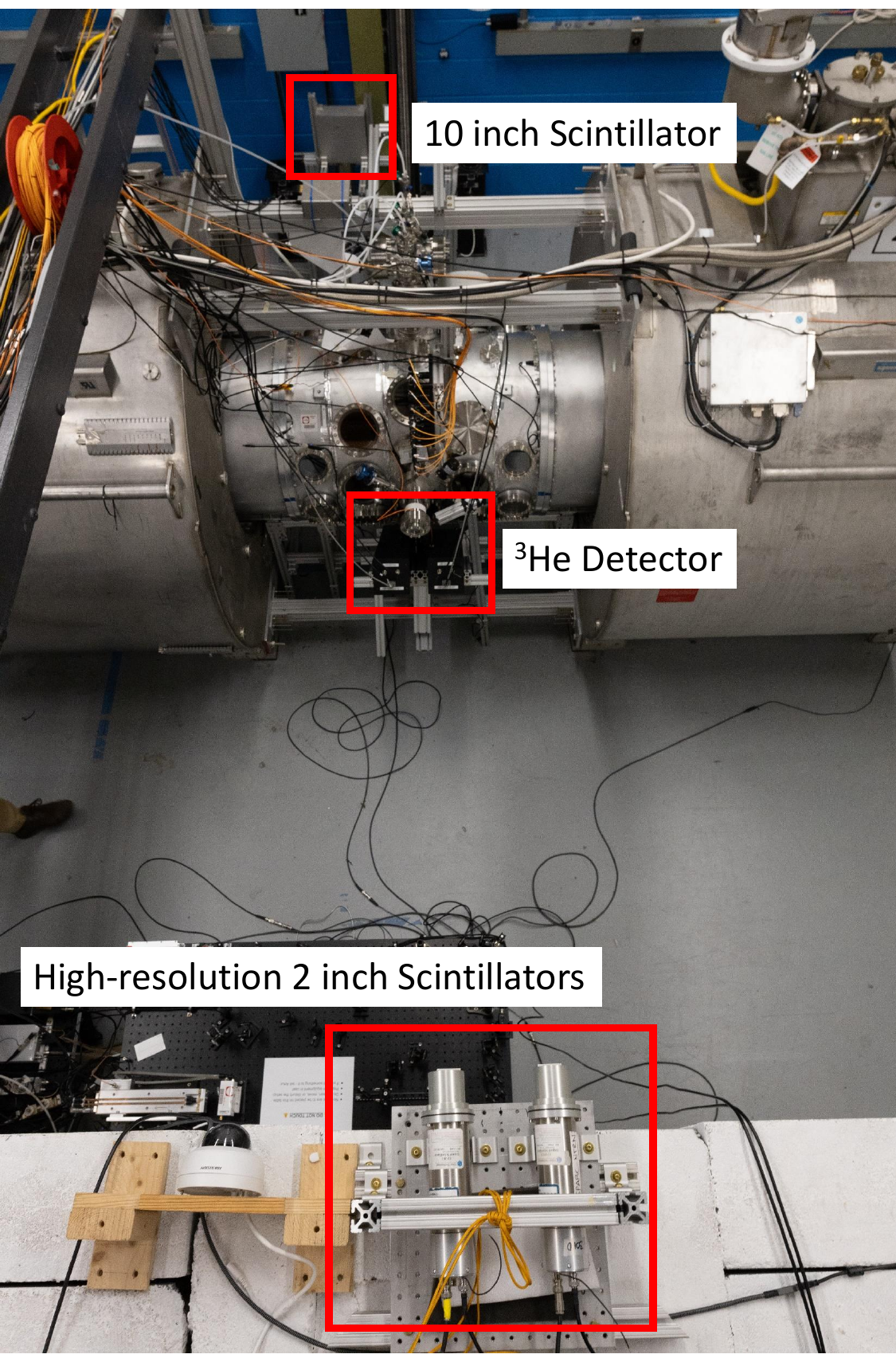}
    \caption{Detector configuration photographed from top of radiation shield wall. From top of the image to bottom: the 10-inch EJ-301 detector is placed 2.6 m from the machine center point, balancing sensitivity to neutron emissions with the magnetic shielding capabilities of its photomultiplier tube; $^3$He detectors fixed to the vacuum vessel, maximizing sensitivity to neutron emissions; two high-resolution EJ-301/D detectors placed 3.1 m away from the machine center on top of the shield wall, minimizing the gain reduction caused by the stray magnetic field.}
    \label{fig:photo}
\end{figure}

The neutron production of CMFX plasmas is studied with a set of complementary detectors to determine the total fusion rate: two 2-inch by 2-inch right cylindrical EJ301/D liquid organic scintillators are used to make \textit{in silico} calibrated measurements of the total fusion neutron yield for a set of high performing discharges; a 10-inch by 10-inch by 3-inch square prism EJ301 detector was then cross-calibrated to these two smaller detectors and used to measure lower yields and investigate the time evolution of the fusion rate; and a permanently installed $^3$He detector was calibrated \textit{in situ} using a $^{252}$Cf neutron source to make independent measurements of the total fusion yield. Figure~\ref{fig:photo} shows the detector layout around the experiment. 

\subsection{Absolutely \textit{in silico} calibrated neutron yield measurements with 2-inch scintillators}
\label{subsec:2in_abs_cal}

\subsubsection{Experimental setup and data processing}
\

Two high-performance detectors were selected for absolute measurements of CMFX neutron yields. Both are 2-inch by 2-inch right cylindrical volumes of xylene-based liquid organic scintillator coupled to Hamamatsu R7724 photomuliplier tubes (PMTs). The scintillating solutions (EJ301 and EJ301D\cite{Becchetti_2016}, sourced from Eljen Technology) of the two detectors are chemically identical and differ only in the isotope of hydrogen present ($^1$H and $^2$H respectively). Deuterated organic scintillators offer enhanced pulse shape discrimination (PSD) and spectrometric properties compared to standard hydrogen-based scintillators, at the expense of reduced light output due to the kinematics of neutron scattering on deuterium. Deuterated-xylene is being investigated for use on future fusion experiments. The pulse shape discrimination (PSD) capabilities of these detectors have previously been characterized using an Americium-Beryllium neutron source, and both detectors demonstrated a PSD figure of merit greater than 1 over a large pulse height range \cite{Ball2024}.

The substantial stray magnetic field produced by the CMFX magnets complicates the use of PMT-based detectors in the area immediately surrounding the machine. Thus, the 2-inch detectors were mounted 3.1 m from the midpoint of CMFX in a region where the field strength was $\sim$7 Gauss as measured with a Lake Shore Model 460 3-axis Hall probe. In this region, the mu-metal shields of the detectors were sufficient to provide performance comparable to that observed in Earth's background field alone, determined with the use of a 0.6 $\mu$Ci $^{252}$Cf source. A $^{60}$Co source was used to energy calibrate the detectors using the method of Dietze and Klein \cite{dietze_klein}. 

A CAEN DT1471ET high voltage power supply was used to supply both detectors with $-1000$V. The anode output of each unit was then directly connected to a 500 MS/s 14bit CAEN DT5730SB digitizer running the CAEN DPP-PSD firmware. This firmware allows for a short-gate and long-gate integral to be computed in real time with the onboard FPGA for rapid PSD analysis. The short gate was set to 34 ns, the long-gate to 250 ns, and the pregate time to 14 ns. Data from the digitizer were acquired using the CAEN CoMPASS software. Absolute timing information was provided by routing the CMFX discharge timing signal into the DT5730SB digitizer. PSD values for each pulse were computed with the standard ``tail-to-total" ratio:

\begin{equation}
    \text{PSD} = \frac{Q_{\text{long}} - Q_{\text{short}}}{Q_{\text{long}}}
\end{equation}

\begin{figure}[]
    \centering
    \includegraphics[width=\linewidth]{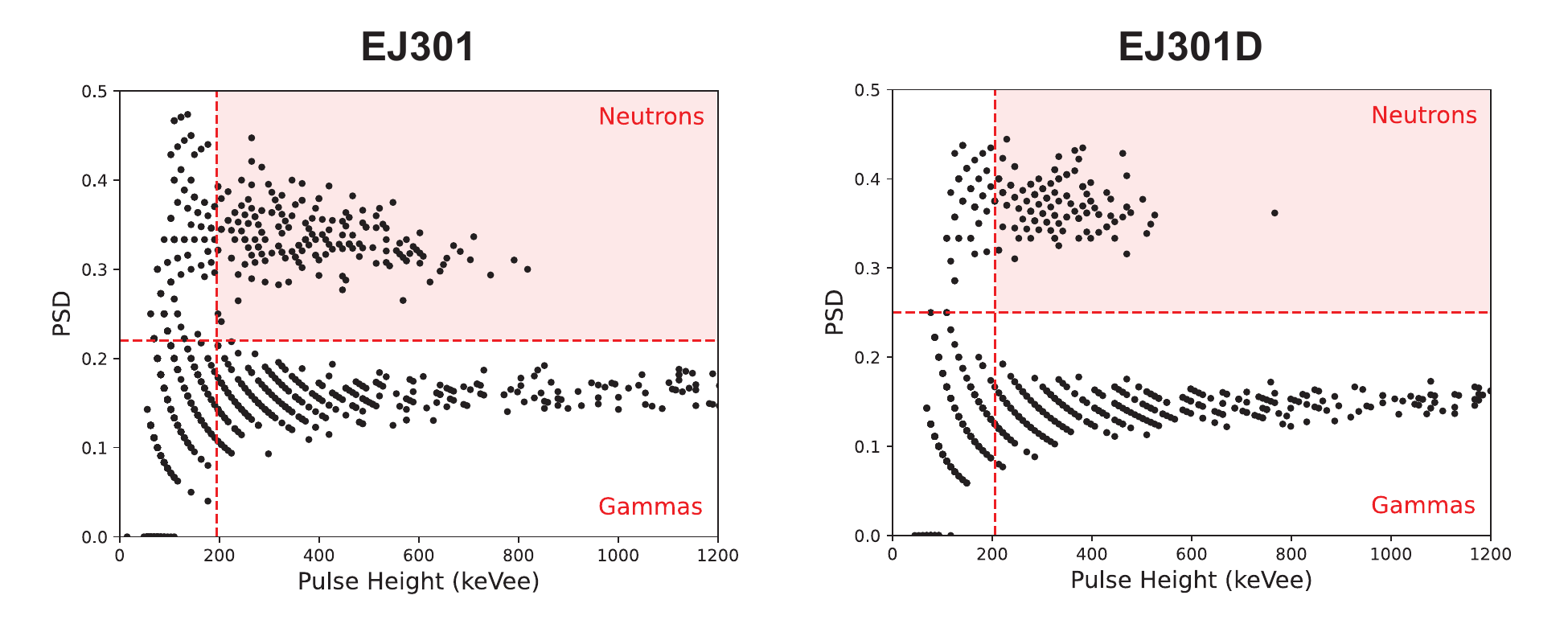}
    \caption{Pulse shape vs. pulse height distributions measured by 2-inch calibration detectors during experiments on CMFX. Left panel shows hydrogen-based EJ301 and right shows deuterium-based EJ301D. These plots are constructed from all data collected over 120 experimental discharges. The neutron (above) and gamma (below) populations are well separated, enabling identification of the neutron induced counts and therefore absolute measurement of the neutron yield. The PSD and pulse height thresholds are indicated with dashed red lines, with the shaded region indicating which pulses were used for the yield analysis.}
    \label{fig:2in_PSD_histograms}
\end{figure}

Where $Q_{\text{long}}$ and $Q_{\text{short}}$ are the charge integral of the pulse over the long and short gates respectively. Fig.~\ref{fig:2in_PSD_histograms} shows a PSD versus pulse height scatter plot for both detectors aggregated from all CMFX discharges analyzed and demonstrates that the PSD performance achieved was equivalent to previous studies using these detectors \cite{Ball2024}. In all subsequent analysis, PSD thresholds of 0.22 and 0.25 were used for the EJ301 and EJ301D detectors respectively, and a pulse height threshold of $\sim$ 200 keVee was implemented in both detectors to reject the region of pulse height space where PSD performance is degraded. The resulting discrimination of neutron and gamma signals allows the neutron response of these detectors to be completely isolated and enables accurate modeling of the detectors' coupling to CMFX neutron emissions.

Due to the relatively low total efficiency of these detectors, we consider shot-integrated neutron yields to accrue maximum counting statistics. We consider only the counts acquired during the plasma's flat top, defined as the time period for which the magnitude of the applied bias voltage is above 95\% of its set value. See Fig. \ref{fig:60kV shots} in Sec. \ref{sec:performance} for time traces from a typical CMFX discharge. We define the average detector count rate to be the number of counts that occur above threshold during the plasma flattop divided by the flattop time. Next, we compute the total efficiency of these detectors in order to relate this average count rate to the average total neutron yield rate of the device.

\subsubsection{Computational modeling of total detector efficiency}
\

While performing neutron yield measurements with a pulse counting detector are conceptually straightforward, in reality determining the coupling of measured counts to source intensity demands an intimate understanding of the subtleties of neutron transport. In order to compute $\epsilon$, defined in Eq.~\ref{yield_eq}, one must leverage detailed knowledge of the source emission distribution in energy, angle, and space, as well as the geometry and composition of surrounding materials and the detector response. In this work, we compute the total efficiency of each 2-inch detector by factoring it into a total zero-threshold efficiency, $G$ and a finite-threshold detection efficiency, $T$:

\begin{equation}
    Y_n=\frac{M}{\epsilon} = \frac{M}{G T}
\end{equation}

\begin{figure}
    \centering
    \includegraphics[width=0.9\linewidth]{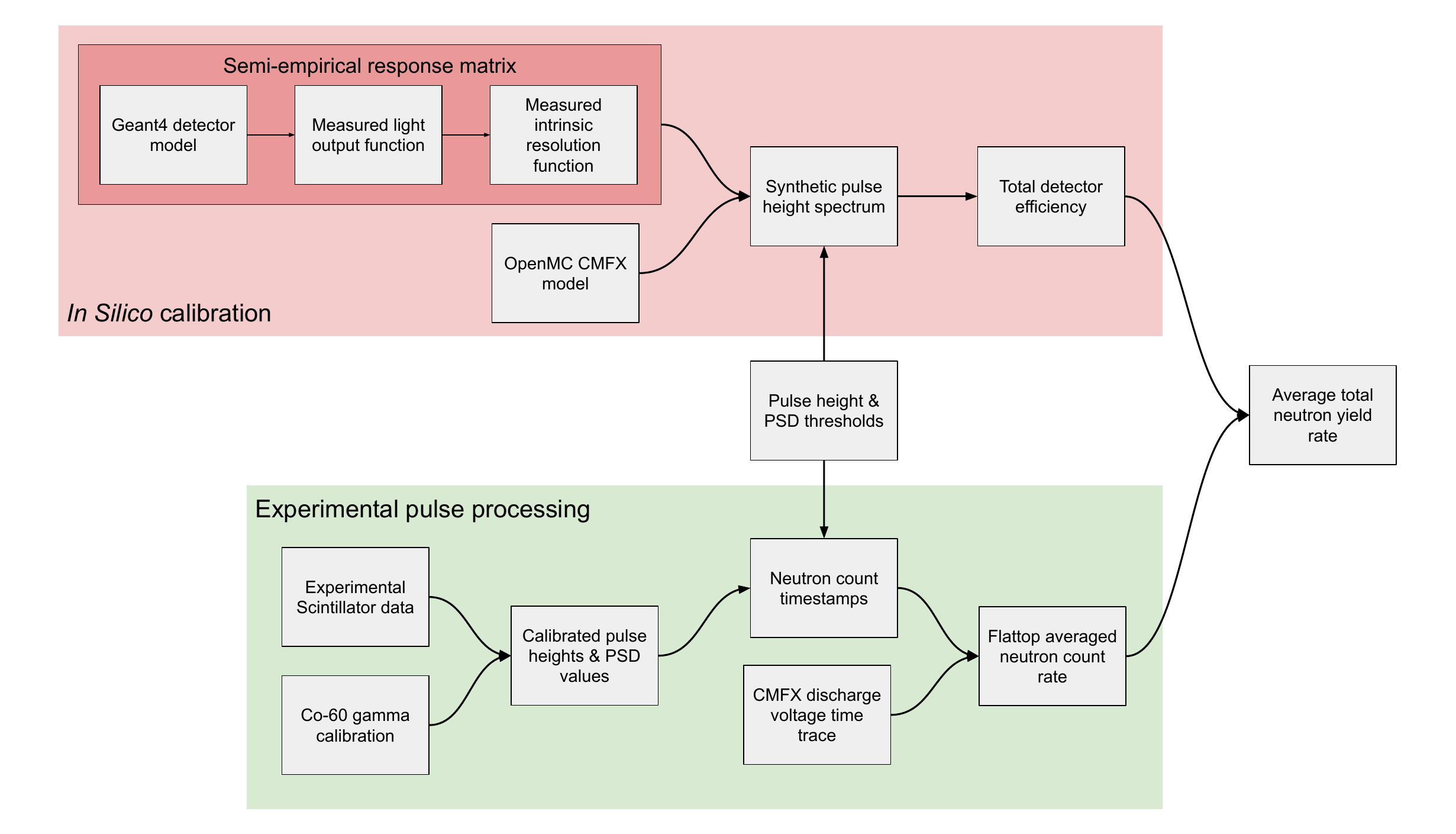}
    \caption{A flow chart outlining the absolute calibration and analysis workflow used for the 2-inch liquid scintillators in this work. The highlighted red region outlines the \textit{in silico} calibration, based primarily on coupled detector response and neutron transport simulations. The green region outlines the processing performed on the collected experimental pulse data, which is then converted into an absolute yield rate with the efficiency computed by the \textit{in silico} calibration.}
    \label{fig:LOS_flow_chart}
\end{figure}

The total zero-threshold efficiency is defined as the fraction of emitted source neutrons which elastically scatter within a given detector, and the finite-threshold detection efficiency is defined as the fraction of elastic scattering events which generate a measured count above both pulse height and PSD thresholds. We compute each of these factors with coupled simulations: $G$ is assessed using a simplified CMFX model built in the Monte Carlo neutronics code OpenMC \cite{ROMANO2015}, and $T$ is calculated using a Geant4 \cite{geant4} model combined with post-processing to incorporate the effects of scintillation and photon detection. The OpenMC model tallies the energy-resolved scattering rate spectrum for both 2-inch scintillators, in units of elastic (n,p) or (n,d)  scattering reactions per source neutron. This spectrum is then multiplied by the semi-empirical detector response matrix generated using the Geant4 model to produce a synthetic detector pulse height spectrum. From this synthetic response, the total efficiency above a given threshold can be computed by summing all bins above that threshold. This sum is equal to the above-threshold efficiency because the response matrix is normalized such that the integral of each pulse height row is 1, ensuring that applying the response matrix to a given scattering rate vector maps each tallied scatter to a single synthetic pulse. A flow chart outlining the procedure used to compute the total detector efficiency and process experimental pulse data to calculate the average yield rate can be found in Fig.~\ref{fig:LOS_flow_chart}.

\begin{figure}[H]
    \centering
    \includegraphics[width=\linewidth]{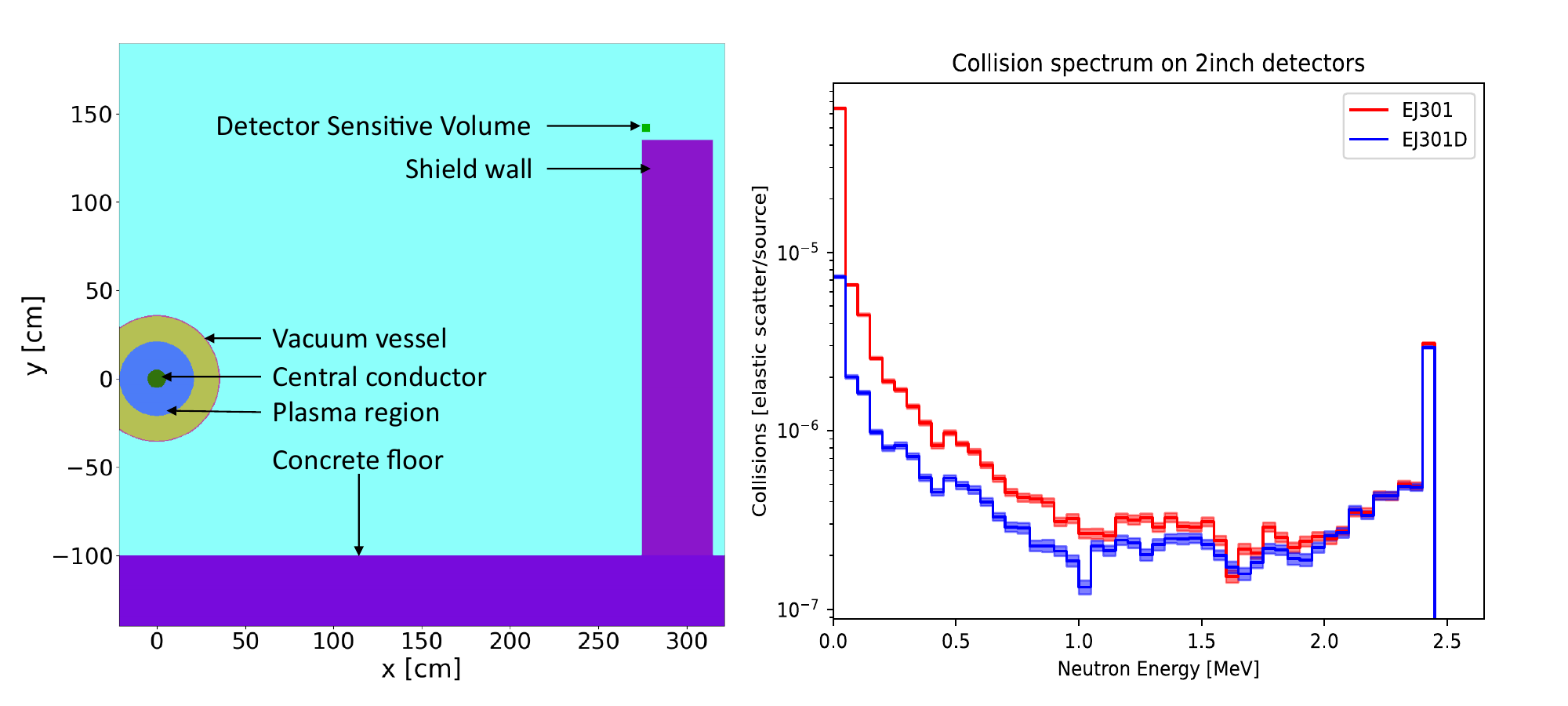}
    \caption{OpenMC transport model and scattering rate spectra. The left figure shows a vertical plane of the geometry used in the neutron transport simulation. The detector sensitive volume is shown as the small green square atop the concrete shield wall shown in purple. The right figure shows the computed neutron scattering spectrum with protons or deuterons EJ301 and EJ301D detectors, respectively. The shaded region around the spectra indicate the uncertainty from Monte Carlo statistics. The difference in scattering cross section for protons and deuterons manifests in the downscattered collision spectrum.}
    \label{fig:openmcSpectrum}
\end{figure}

The OpenMC model implements a centrifugal mirror plasma neutron source as well as major structures influencing neutron transport, including the magnets, vacuum vessel, and radiation shield wall; more details on the OpenMC model may be found in Schwartz's doctoral thesis \cite{NickPhDthesis}. A vertical plane of the model and the computed neutron collision spectra are shown in Figure \ref{fig:openmcSpectrum}. Integrating these flux spectra and multiplying by the detector sensitive area gives zero-threshold total efficiencies of $G_{301}=(1.021\pm0.007)\times 10^{-4}$ and $G_{301D}=(2.92\pm0.02)\times 10^{-5}$ elastic collisions per emitted neutron. The quoted uncertainties correspond to Monte Carlo statistics only. Because these uncertainties are both $< 1\%$, they are neglected for the remainder of the analysis.

\begin{figure}
    \centering
    \includegraphics[width=\linewidth]{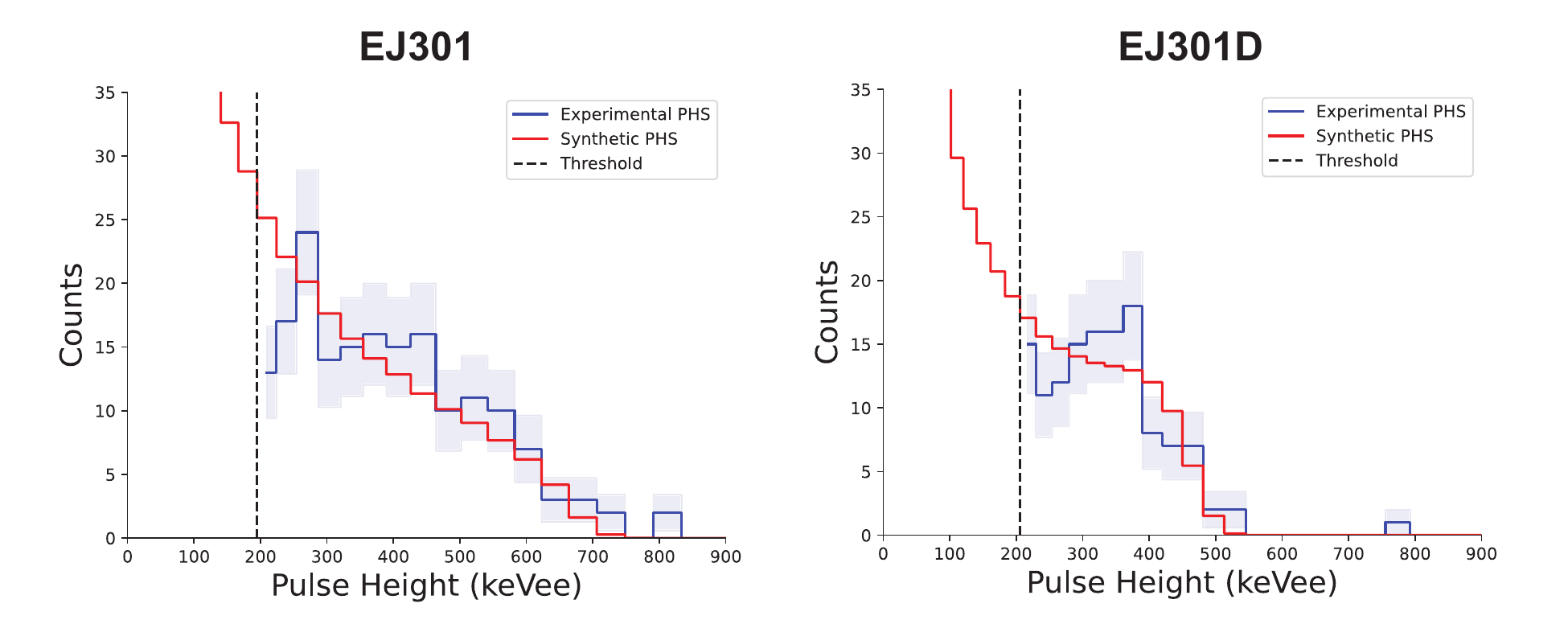}
    \caption{Plots of cumulative experimental pulse height spectra over all discharges used in this study along with the synthetic pulse height spectra produced by the \textit{in silico} calibration method described in Sec. \ref{subsec:2in_abs_cal}. The $^1$H based EJ301 detector is plotted on the left, and the $^2$H EJ301D detector on the right. The synthetic pulse height spectrum is scaled such that its integral above threshold is equivalent to that of the experimental pulse height spectrum.}
    \label{fig:2in_cumulative_phs}
\end{figure}

The detector response matrices are calculated using the Geant4-based grasshopper \cite{grasshopper} code, which captures neutron scattering and ion slowing down physics. The output of this simulation is then combined with experimentally measured light output and intrinsic resolution functions to model the impact of scintillation and photon detection on the detector response. We refer to this approach to response matrix generation as ``semi-empirical" because of its combination of simulation output and empirical measurement.  The generation of these response matrices is discussed in greater depth in Ref.~\cite{Ball2024}. These matrices are used to compute synthetic pulse height spectra from the simulated scattering rate spectrum. As discussed above, this matrix is normalized such that the total scattering rate is preserved upon application of the matrix. This allows for a tallied scattering rate spectrum to be transformed into a counts-per-source-neutron pulse height spectrum, where the magnitude of each bin is the probability of a source neutron generating a pulse of that magnitude in the detector. We compute the above-threshold efficiency for these detectors on CMFX to be $T_{\text{301}}$ = $0.0389$ and $T_{\text{301D}}$ = $0.122$. The uncertainty in the gamma calibration, which effects the scaling of the pulse height spectrum with respect to the threshold, was found to be $\sim 1\%$, and thus was neglected. This is supported by the strong agreement found between synthetic and experimental pulse height spectra, shown in Fig.~\ref{fig:2in_cumulative_phs}. When combined with the total zero-threshold efficiency, this gives a total above-threshold efficiency of  $\epsilon_{\mathrm{301}}$ = $3.97 \times 10^{-6}$ and $\epsilon_{\mathrm{301D}}$ = $3.56 \times 10^{-6}$ counts per source neutron. 

This modeled total efficiency can then be used to convert an experimentally measured average count rate into an absolute neutron yield rate. It is important that the simulated threshold is accurately matched to the experimentally imposed threshold for signal counts. Figure \ref{fig:2in_cumulative_phs}
plots the cumulative experimental pulse height spectra for each 2-inch detector over all shots used in this analysis as well as the synthetic pulse height spectrum generated by the efficiency calculation described above. We find that the synthetic spectra match the experimental spectra well, with a reduced chi-square of 1.57 for EJ-301 and 1.04 for EJ-301D. The simulated pulse height spectra were scaled such that the integrals of the experimental and simulated spectra above the pulse height threshold are equal, thus allowing comparison of pulse height spectrum shape alone.

A challenge in the application of this approach to total yield measurements is that estimating bias and uncertainty introduced by the computational models is nontrivial. While OpenMC and Geant4 have been extensively benchmarked \cite{Ebiwonjumi_2025, Segantin_2024, Bae_2022, ALLISON_2016}, the accuracy of these codes is dependent on the quality of the inputs provided. An effort was made to make the model as accurate as possible based on known dimensions and materials used in the construction of CMFX and the shield wall upon which the 2-inch detectors were placed. For the purposes of this study, where Poisson uncertainty dominates ($\sim 20 - 30$\%), we choose to neglect this contribution to the total uncertainty. This choice is further motivated by the following: the signal is dominated by direct neutrons, limiting the impact of geometric and materials uncertainties, and because these response matrices have been shown to agree well with measured detector responses (see Fig. \ref{fig:2in_cumulative_phs}). Our choice to neglect these uncertainties is supported by the strong agreement we find between this method and the \textit{in situ} calibrated $^3$He tubes, discussed in detail in the following section and shown in Fig.~\ref{fig:Scint-3He comparison}. Nonetheless, future work extending this methodology should look to better quantify these sources of model uncertainty. For all remaining analyses, only Poisson uncertainty is considered for yield measurements made with the 2-inch scintillators alone. 

\subsection{Cross-calibrated neutron emission rate measurements with 10-inch scintillator}

To complement the relatively insensitive absolute calibration scintillators, a more sensitive scintillator was cross-calibrated to measure low yields and the time resolved emission rate for high yield experiments. This detector is a 10-inch by 10-inch by 3-inch square prism of EJ301 coupled to a 3-inch PMT and has previously been studied by Hartwig et al. \cite{HARTWIG2014}. 

\begin{figure}[H]
    \centering
    \includegraphics[width=\linewidth]{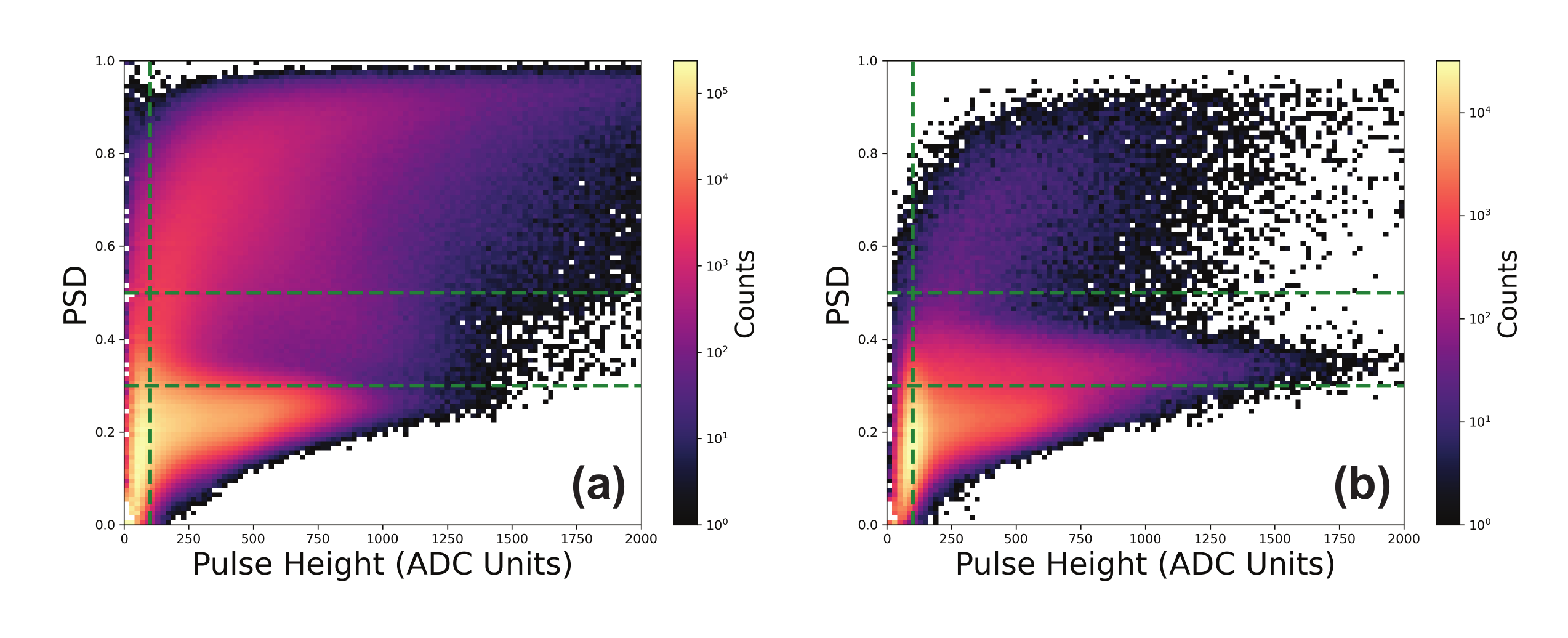}
    \caption{2D PSD histograms measured by the 10-inch detector for background gammas (a) and $^{252}$Cf (b). Neutron counts manifest as a population above PSD values of 0.3, indicated by the dashed line. The population at high PSD parameter is due to pulse pile-up. Since the neutron-gamma separation is much worse compared to the 2-inch detectors, cross calibration is required to relate the measured count rate to the neutron emission rate. The background gamma plot has a larger number of total counts because data were taken over a longer time interval than in the $^{252}$Cf data.}
    \label{fig:10inchCheckSources}
\end{figure}

Figure \ref{fig:10inchCheckSources} shows the 2D PSD histograms measured by this detector when exposed to a $^{252}$Cf neutron source, placed directly in front of the detector, and background gammas from naturally occurring sources in the CMFX experiment hall. While the $^{252}$Cf plot clearly shows a neutron distribution which is not present in the background measurement, it is also clear that it is not well separated from the gamma distribution below. Three thresholds are applied to reject as many non-neutron counts as possible: a lower PSD threshold of 0.3, an upper PSD threshold of 0.5, and a pulse height threshold of 100 ADC units. Counts above the upper PSD threshold are the result of pile-up events. The inability of this detector to clearly discriminate between gamma and neutron counts rules out the absolute calibration method described in Section \ref{subsec:2in_abs_cal}. Instead, we cross-calibrate this detector to the yield measured by the 2-inch detectors.

The increased sensitivity and poor PSD performance of this detector necessitated background subtraction to isolate the signal produced by CMFX, as background gammas were not completely rejected by PSD thresholds. The average background count rate within the neutron PSD region, $\mathrm{\dot M_{B, 10in}}$ was calculated from data acquired for $\sim$ 3 seconds before the plasma was initiated. The signal rate $\mathrm{\dot M_{S,10\text{in}}}$ is then found by subtracting this average background rate from the total count rate $\dot M_{10in}$.
\begin{equation}
    \mathrm{\dot M_{S,10in} = \dot M_{10in} - \dot M_{B, 10in}}
\end{equation}

A statistically significant number of counts in the 2-inch detectors were accrued over several repeated discharges at four different voltages to cross-calibrate the 10-inch detector to the total neutron emission rate inferred by the \textit{in silico} calibration. We enforce the physically-motivated constraint that the fit pass through the origin and perform an orthogonal distance regression in order to incorporate both the Poisson uncertainty of the yield measurements from the 2-inch detectors and the Poisson statistics of the measured 10-inch count rate into the fit. The constant of proportionality is used as a cross-calibration factor $m_{XC}=\epsilon_{XC}^{-1}=(2.822\pm 189)\times10^3$ neutrons/count to relate the 10-inch signal count rate $\dot M_{S, 10in}$ to fusion yield rate. 

\begin{figure}[H]
    \centering
    \includegraphics[width=0.75\linewidth]{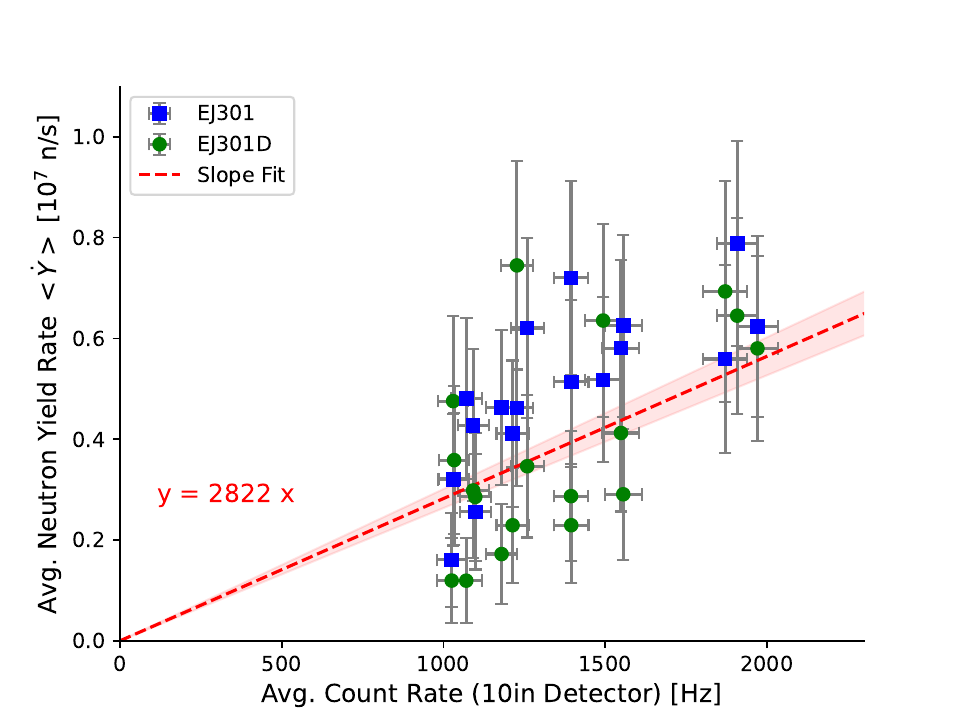}
    \caption{Cross-calibration of 10-inch EJ301 detector count rate with neutron rate measured by absolutely calibrated 2-inch detectors. Blue squares indicate neutron emission rate inferred from EJ301 data, while green circles represent EJ301D data. Vertical and horizontal error bars correspond to Poisson uncertainty in the 2-inch and 10-inch detectors respectively. The dashed red line plots the slope fit to the data used to convert the 10-inch count rate to a total neutron yield rate. The orthogonal distance regression takes both uncertainties into account in the determination of the cross calibration uncertainty, which is shown as the shaded red region above and below the fit line.}
    \label{fig:10inchCrossCalib}
\end{figure}

Fig. \ref{fig:10inchCrossCalib} shows the linear fit used to extract the cross-calibration factor for the 10-inch detector. The yield measurements made by each of the 2-inch detectors are shown in blue and green for EJ301 and EJ301D respectively. We find that the two populations show strong overlap within error bars, indicating little to no systematic bias between the two detectors. We also observe a strong linear trend in the data, as expected. The linearity of this trend is supported by the residual chi-square of the fit, which is computed to be 0.98. For the remainder of this work, all scintillator yield measurements presented are made with this cross-calibrated 10-inch detector and the corresponding uncertainty is the Poisson and calibration slope uncertainties summed in quadrature. This cross-calibration technique allows for the larger set of counting statistics accrued with the 2-inch detectors over many discharges to be combined in such a way that a larger chunk of the CMFX yield range can be resolved with a lower total uncertainty. 

\begin{equation}
    \dot Y_n = m_{XC}\dot M_{S, 10in}
\end{equation}

\subsection{Absolutely calibrated neutron yield measurements with $^3$He proportional counters}
Two $^3$He gas proportional neutron counters are permanently installed at CMFX, although only one is used in this study due to an electrical fault in the other. These detectors have a right cylindrical gas chamber 1-inch in diameter and 8-inches in length and are pressurized with $^3$He to 4 atm. The gas cylinders are encased in a 4-inch shell of high density polyethylene. A detection event occurs when an incident neutron undergoes a nuclear reaction with the $^3$He, producing a triton and a proton and liberating 765 keV of energy \cite{Henzlova2024}.

\begin{equation}
    \mathrm{^3He + n \to~^3H + ~^1H + 765 \ keV}
\end{equation}

The total efficiency of this detector was determined via an \textit{in situ} calibration procedure using a $^{252}$Cf neutron source. This source was manufactured at Oak Ridge National Laboratory in 1970 and calibrated by NIST in 1985. The intensity, $S_0$, of this source was determined to be 15510$\pm775$ n/s as of March 4th, 2025. $^{252}$Cf was selected because its peak neutron emission is at $\sim2.5$ MeV, making it a reasonable analog to 2.45 MeV DD fusion neutrons. However, $^{252}$Cf is not a monoenergetic source and emits neutrons over a wide range of energies. In this analysis we neglect any effects resulting from the difference in neutron spectrum between calibration and operation and assume the $^3$He detector responds identically to $^{252}$Cf neutrons and to DD fusion neutrons.

In order to relate measurements of a point source to an extended source, we represent equation \ref{yield_eq} as an integral of the product of the emissivity density, $S(\vec x)$, with the position-dependent weight function, $W(\vec x)$, over the plasma volume.
\begin{equation}
\label{weight_fun_eq}
    \dot M_{\text{He-3}} = \epsilon_{\text{He-3}} \dot Y_n = \int_{V_p}S(\vec x) W(\vec x)d^3x
\end{equation}
$W(\vec x)$ relates the detector's response to neutrons emitted from a given point in space. $W$ is a function of the geometry and materials of the experimental setup and is independent of the source. Noting that the total yield rate is just the volume integral of the emissivity density, we can rearrange Eq.~\ref{weight_fun_eq} to get an expression for the total detector efficiency:
\begin{equation}
    \epsilon_{\text{He-3}} = \frac{\int_{V_p}S(\vec x) W(\vec x)d^3x}{\int_{V_p}S(\vec x) d^3x}=\langle W(\vec x)\rangle
\end{equation}
Thus, the detector's coupling constant to a given source is the source-averaged weight function. The weight function at any point, $\vec x_i$, can be empirically measured using a point source at that point with emissivity given by $S(\vec x) = S_0\delta(\vec x -\vec x_i)$:
\begin{equation}
    \dot M_{\text{He-3}}(\vec x_i) = \dot M_i = \int S_0\delta(\vec x - \vec x_i)W(\vec x)d^3x = S_0W(\vec x_i)
\end{equation}
\begin{equation}
    W(\vec x_i) =W_i =  \frac{\dot M_i}{S_0}
\end{equation}

We adopt the simple model that the CMFX emissivity density is represented by an azimuthally symmetric ring source with radius $R_0$ on the midplane, reducing the volume integration to a single integral over $\theta$. The integral can be approximated by a finite sum over a set of $N$ measurements of the weight function spaced by $\Delta\theta_i$, where $\sum_i ^N{\Delta\theta_i}=2\pi$:
\begin{equation}
    \epsilon_{\text{He-3}} =\frac{\int_{V_p}S_0\delta(r-R_0)\delta(z) W(\vec x)d^3x}{\int_{V_p}S_0\delta(r-R_0)\delta(z) d^3x}= \frac{1}{2\pi}\int_0^{2\pi}W(\theta)d\theta= \frac{1}{2\pi}\sum_i^NW_i\Delta\theta_i
\end{equation}
The calibration uncertainty is computed by summing the uncertainties of intermediate quantities in quadrature, as the errors are uncorrelated:
\begin{equation}
\label{He3_uncertainty}
    \frac{\delta \epsilon_{\text{He-3}}}{\epsilon_{\text{He-3}}}=\sqrt{\sum_i^N\left[ \left( \frac{\delta M_i}{M_i}\right)^2+\left( \frac{\delta \Delta\theta_i}{\Delta\theta_i}\right)^2\right]+\left( \frac{\delta S_0}{S_0}\right)^2}
\end{equation}

\begin{figure}
    \centering
    \includegraphics[width=0.75\linewidth]{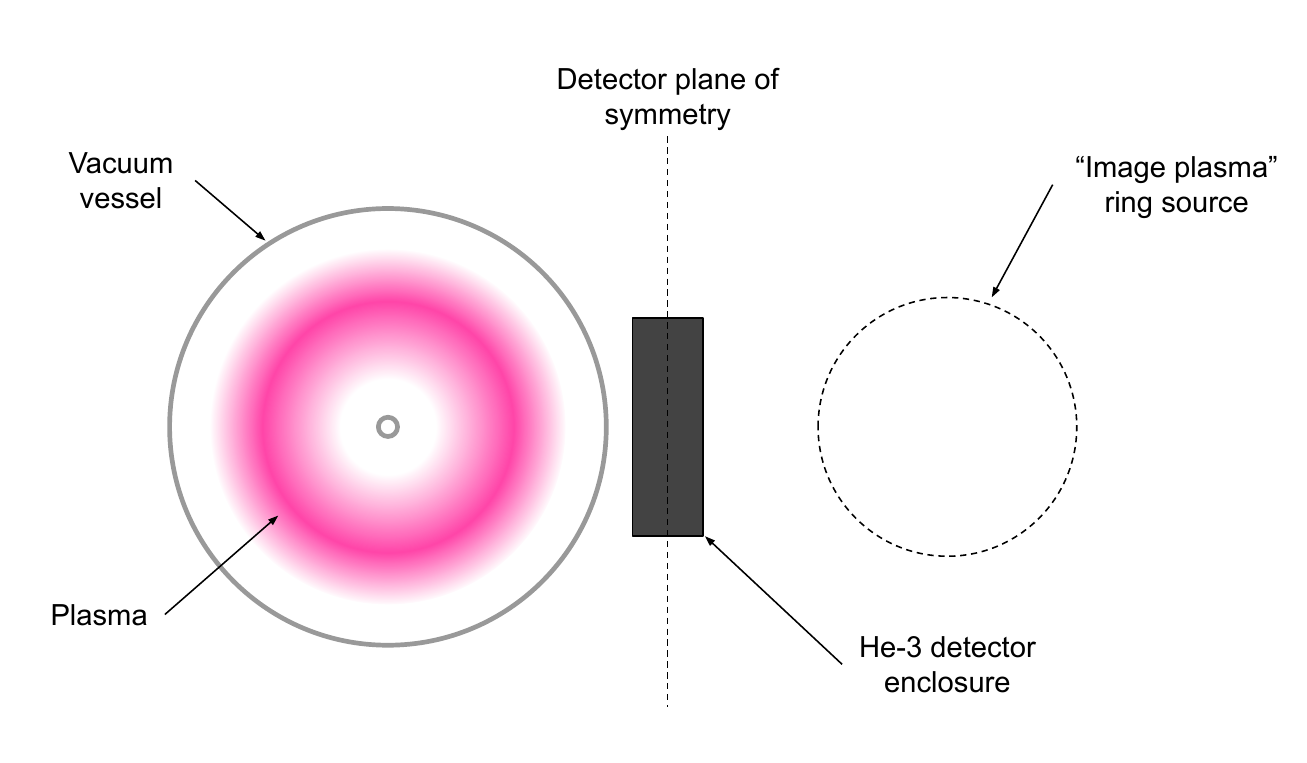}
    \caption{Schematic diagram of mirror ring source geometry used to calibrate $^3$He tubes without requiring the CMFX vacuum vessel to be vented. The symmetry of the detector response is exploited, allowing the calibration source to be placed at locations outside of the vacuum vessel with the same geometric efficiency as those inside the plasma. This technique assumes that attenuation from the vacuum vessel can be neglected.}
    \label{fig:He-3_calib}
\end{figure}

To avoid venting CMFX, we leveraged the symmetry of the detector geometry with respect to the plasma and defined an equivalent `image plasma' region by reflecting about the detector plane. The calbiration source can then be placed in this `image plasma' region without interrupting the vacuum or biasing the calibration significantly. This assumes that neutron interactions with the vacuum vessel contribute negligibly to the detector response. This assumption is supported by the fact that the CMFX vacuum vessel is made of 0.25-inch (0.64 cm) thick aluminum, in which the mean free path of 2.45 MeV neutrons is $\sim6-7$ cm.  See Fig. \ref{fig:He-3_calib} for a schematic diagram of the image ring source used to locate the calibration source. A calibration structure was constructed to position the $^{252}$Cf source at various angular positions mirroring the peak emission region of the plasma. We placed the source at 6 angular locations in the upper half-arc arc and exploited the up-down symmetry of the geometry to assert that $W(\theta)$=$W(-\theta)$. The $^3$He detector's count rate was measured in each position with counting uncertainty $\delta M_i/M_i < 1.5\%$, and the position of the source was recorded with 5~mm precision which yields an uncertainty in the angular spacing $\delta\Delta\theta_i/\Delta \theta$ ranging from 2-12\%. Using this method, we estimate the total efficiency of the $^3$He proportional counters to be $\epsilon=(6.2 \pm 1.4) \times 10^{-4}$ counts per source neutron. This estimate agrees well with the neutron absorption rate per source neutron in $^3$He tallied by the OpenMC model which is $(6.02\pm0.01)\times10^{-4}$ absorptions per source neutron. Notably, the OpenMC model includes the aluminum vacuum vessel neglected during the \textit{in situ} calibration procedure.

\begin{figure}[H]
    \centering
    \includegraphics[width=0.55\linewidth]{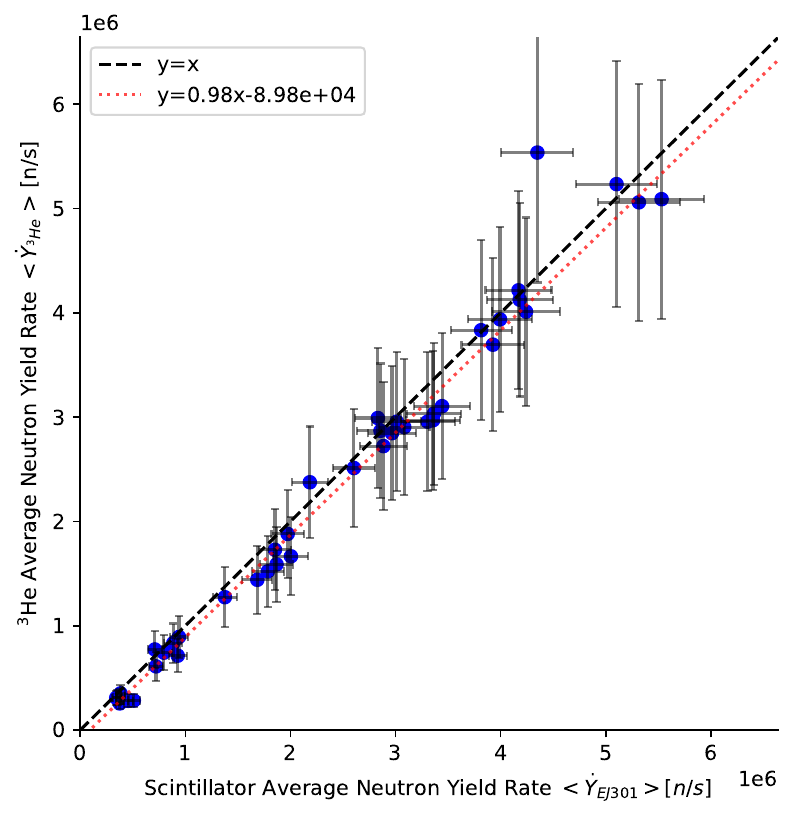}
    \caption{Comparison of \textit{in silico} calibration of 10-inch EJ301 detector and \textit{in situ} calibration of $^3$He tube detector using data collected over 42 discharges. Scintillator error bars are calculated using Poisson statistics which dominate the uncertainty. $^3$He error bars are calculated using Eq.~\ref{He3_uncertainty}. The dashed black line indicates equality, while the dotted red line shows the result of a regression of the inferred yields. The fact that the slope of the linear regression is near unity and the intercept is small demonstrates agreement between yield rates inferred by these two independent methods.}
\label{fig:Scint-3He comparison}
\end{figure}

The average neutron yield rates measured by the $^3$He detectors and the 10-inch scintillator are plotted against one another in Figure \ref{fig:Scint-3He comparison}. Data from 42 discharges that attained steady bias voltages ranging from 42.5 kV up to 62.5 kV are included in this comparison. An orthogonal distance regression is performed on these data and the resulting linear fit, including both slope and intercept as free parameters, finds a slope of 0.981$\pm$0.024 and intercept of $8.98\times10^4$ ($<2\%$ of max value), indicating the methods are not strongly biased relative to one another. The coefficient of determination for this fit is $\mathrm{R}^2=0.978$. The agreement between the two methods of determining the neutron yield rate builds confidence in these results.

\section{Measurements of Fusion Yield on CMFX}
\label{sec:performance}

Using the set of detectors described above, we can now investigate the fusion yield rate of CMFX in a systematic manner. We begin by studying how fusion yield rate varies in time for several repeated discharges, where we find high consistency across experiments. We then investigate how the yield time dynamics change with three different fueling schemes, which may lead to future optimizations in fueling performance. We then investigate how the average fusion yield rate varies with radial set voltage, where we find an exponential relationship between voltage and average yield rate.

\subsection{Temporal evolution of fusion rate}

\begin{figure}[h]
    \centering
    \includegraphics[width=\linewidth]{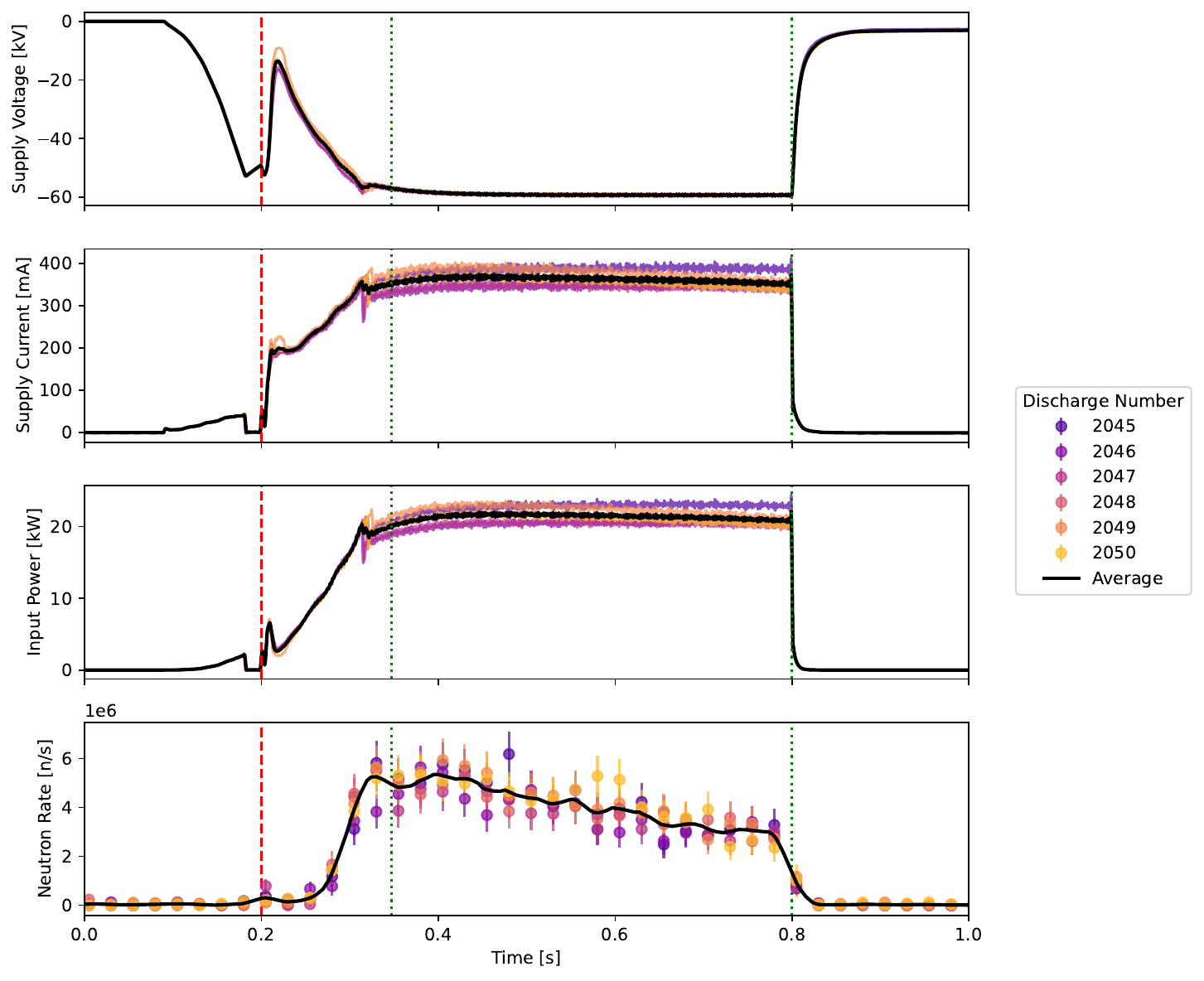}
    \caption{Time traces of measured quantities, including neutron emission rate with 25 ms time resolution by the 10-inch scintillator, from 6 repeated discharges with identical fueling and bias voltage. Vertical red dashed line indicates when the fuel was injected, vertical green dotted lines indicate when the voltage is at flat top. Note the high degree of repeatability of discharges. There is a clear trend of decreasing fusion rate with time.}
    \label{fig:60kV shots}
\end{figure}

The high sensitivity of the 10-inch detector enables time resolved measurements of the fusion rate and gives insight to the evolution of the plasma over the course of a discharge. Figure \ref{fig:60kV shots} shows a series of six repeated discharges with identical conditions: 60 kV bias set voltage and fuel injection of deuterium gas for 250~$\mathrm{\mu}$s . 

These time traces provide a good overview of typical CMFX operations. We observe three main phases: first, the power supply voltage is ramped close to its set point, at which time fuel is puffed in. This causes the voltage to drop and the power draw to increase as the gas is ionized. The voltage and power then begin to grow as the plasma spins up, increasing the cross-field impedance. Once the voltage reaches its set point, the plasma enters the second stage: flat top operation. During this period, the voltage and power draw are approximately constant. However, we observe that the neutron yield rate drops in time approximately linearly, possibly the result of the density dropping in time. Then, the discharge is safely terminated, marking the third and final stage. We find that CMFX has excellent shot repeatability in all parameters plotted, including neutron yield as measured by the cross-calibrated 10-inch liquid scintillator. Error bars represent Poisson uncertainty only. Note that the power supply could sustain the discharge indefinitely, but the discharge is terminated to avoid sputtering and melting of electrodes and insulators since, at present, they are not actively cooled.

\begin{figure}[h]
    \centering
    \includegraphics[width=\linewidth]{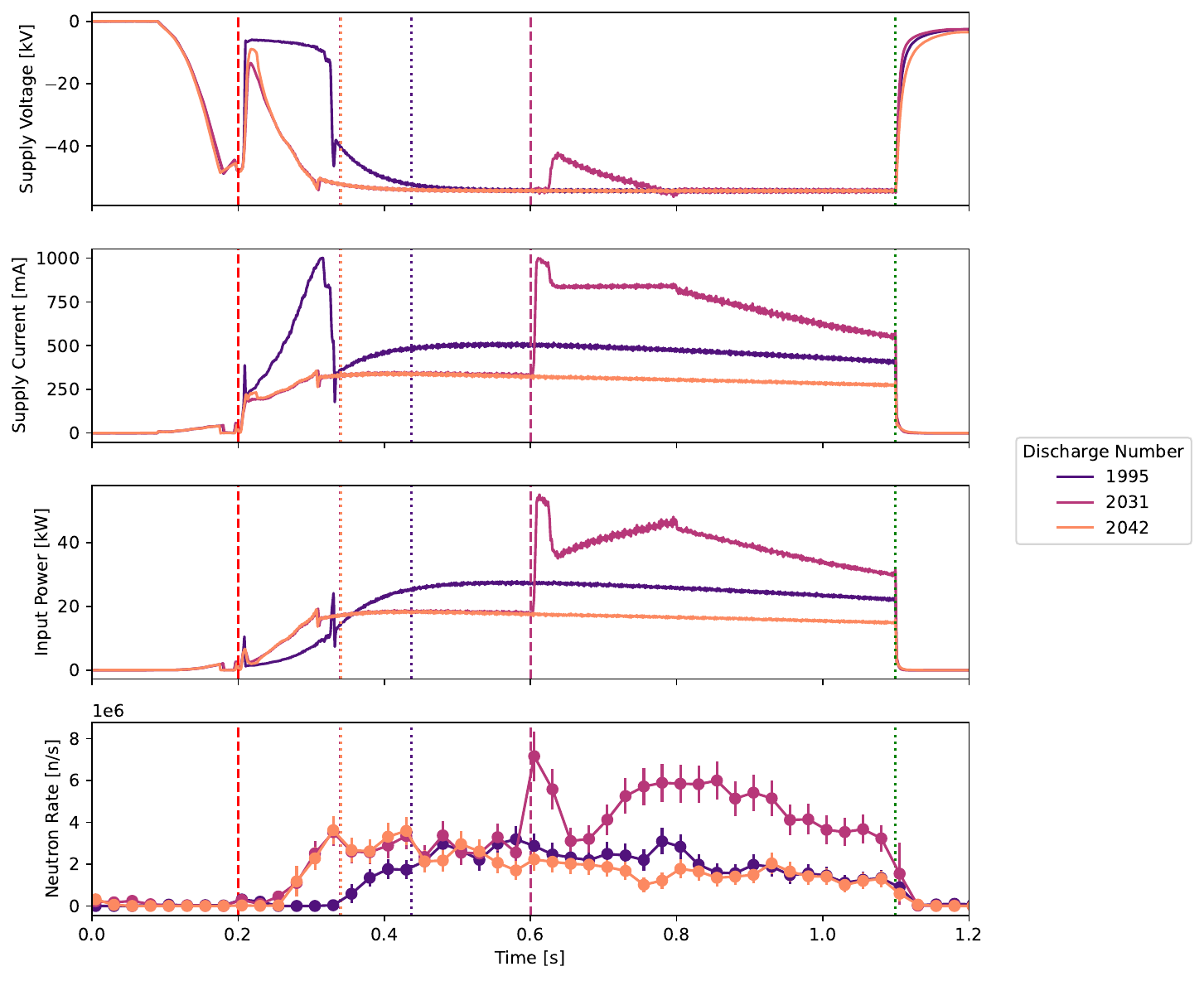}
    \caption{Comparison of different fueling schemes. Discharge 1995, in purple, was fueled by a 1 ms long deuterium gas puff at t=0.2 s. Discharge 2031, in maroon, was fueled by two 0.25 ms gas puffs, the first at t=0.2 s and the second at t=0.6 s. Discharge 2042, in orange, was fueled by a single 0.25 ms gas puff at t=0.2 s. Fueling times are indicated by vertical dashed lines, while dotted lines indicate the beginning of the voltage flattop interval determined as the time when the voltage reaches 95\% the set voltage. The green vertical dotted line indicates the programmed discharge end at t=1.1 s. For all three shots, the set voltage was 55 kV.}
    \label{fig:fueling-study}
\end{figure}

CMFX fueling parameters were also varied to study their impact on machine performance. Three puffing schemes were studied: one puff lasting 1 ms (Discharge 1995), one puff lasting 0.25 ms (Discharge 2042), and two puffs lasting 0.25 ms each spaced by 400 ms (Discharge 2031). Thus, we examine three different total puff durations (0.25~ms, 0.5~ms, and 1~ms) distributed in time either as a single puff at the beginning of the discharge (0.25~ms, 1~ms) or as two puffs separated in time (0.5~ms). Because the fueling rate is constant, total puff-time corresponds linearly to total gas inventory injected. Fig.~\ref{fig:fueling-study} shows the results of each fueling scheme. Comparing the two single-puff scenarios, we find that it takes significantly longer for the 1~ms puff discharge to reach its set voltage, indicating it takes the machine longer to ionize and spin up the plasma, a result of the larger initial gas inventory. Furthermore, the longer puff scenario draws significantly more power than the shorter puff scenario. However, we find that the neutron yield rate is approximately the same for both single-puff scenarios.

A more dramatic difference is observed between the single-puff and double-puff experiments. For the first half of the discharge the 0.25 ms single puff and double puff experiments are nearly identical, as expected based on the results presented in Fig.~\ref{fig:60kV shots}. However, the two experiments rapidly diverge after the second puff is fired at t = 600 ms. The power draw rapidly spikes, with the current draw hitting the supply maximum of 1 Amp. Shortly after this the power supply drops the voltage to reduce the current draw below its maximum. The voltage is then slowly raised again until it returns to the set voltage, now with the current below the operating limit. The total power draw of the plasma remains much larger than that of the single-puff experiment for the remainder of the discharge. We observe a strong correlation between the input power and neutron rate for the double puff experiment, with the neutron rate even replicating the immediate spike after the second puff and the slow build up afterwards. It is worth noting that the double puff experiment, which injected half of the total gas injected by the 1 ms single puff experiment, achieved the highest neutron yield rates of any of the fueling schemes. This indicates that carefully tuning of the fueling scheme may be necessary to optimize the machine performance.

\subsection{Fusion rate scaling with applied voltage}

\begin{figure}[h]
    \centering
    \includegraphics[width=0.75\linewidth]{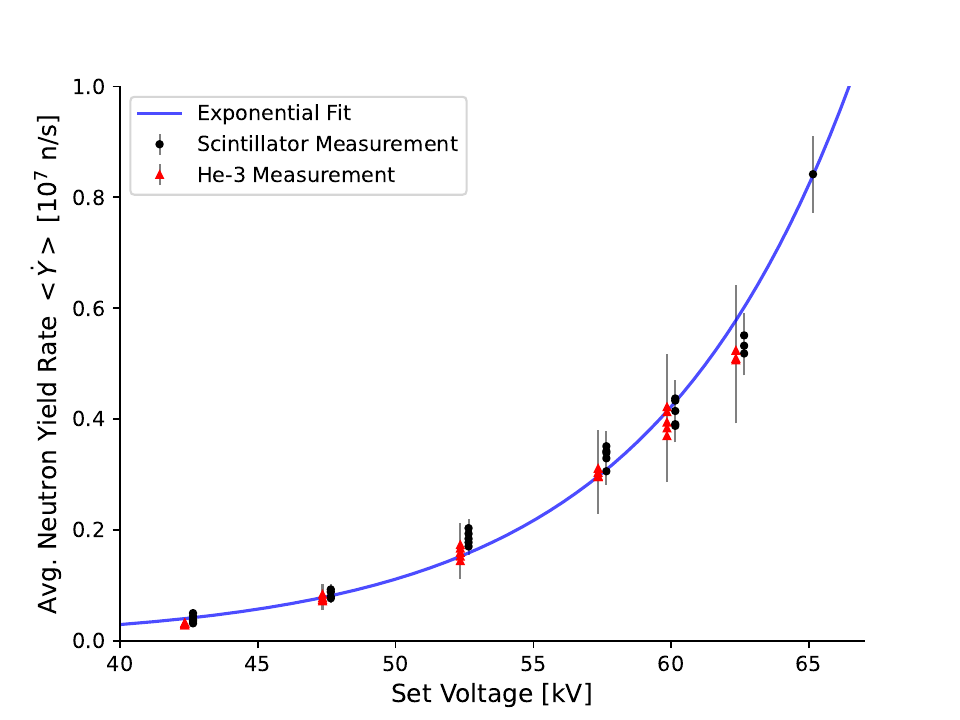}
    \caption{Fusion rate scales approximately exponentially with applied voltage. The functional form of the fitted function is $\dot{Y}(V) = A e^{B(V + C)}$, and was fit to the data with a least squares method. The fitted values of $A$, $B$, and $C$ are 26.0 n/s, 0.133, and -29.9 kV respectively. The exponential dependence of yield on set voltage suggests ion temperature grows linearly with set voltage, as DD reactivity is approximately exponential in temperature for low temperatures. The $^3$He and scintillator measurements were slightly offset in voltage on the plot to improve readability, in reality both sets of points were measured at the same set voltage between the two measurements.}
    \label{fig:voltage_yield}
\end{figure}

The fusion rate as a function of set voltage was studied by measuring the average neutron yield rate rate for 28 discharges with identical fueling schemes at several set voltages. Figure \ref{fig:voltage_yield} shows these data as well as an exponential fit. 

Using a least squares method, the following function was fit to the data:
\begin{equation}
    \dot{Y}(V) = A e^{B (V + C)}
\end{equation}

\noindent Where the fit resulted in A = 26.0 n/s, B = 0.133, and C = $-29.9$ kV. The exponential fit captures the trend in the data well. The good agreement between the data and the exponential fit suggests that ion temperature grows linearly with set voltage, as DD reactivity varies approximately exponentially with ion temperature for low ($\leq$ 1 keV) temperatures \cite{Hutchinson_2002}. This conclusion is further supported by 0D MCTrans++ modeling conducted in the following section.

\section{Physics-Informed Interpretive Modeling with MCTrans++}
\label{sec:modeling}

An interpretive modeling framework was developed using the MCTrans++ code \cite{mctrans} for centrifugal mirror plasma performance to infer plasma parameters that we did not directly measure. The model is briefly introduced, followed by a description of the methods used to combine measurements made with the MCTrans++ tool to extract estimates of ion temperature, electron density, neutral density, and triple product for each shot.

MCTrans++ is a 0D plasma code developed at the University of Maryland to model the unique physics of centrifugal mirrors, including viscous heating, angular momentum confinement, and the centrifugal confining potential. It has been benchmarked against previous centrifugal mirror experiments and has been used to model reactor regimes and parametrically explore operational limits due to charge exchange and Alfv\'{e}n Mach number \cite{mctrans}. In steady state solver mode, MCTrans++ transforms a set of differential equations (species particle continuity, species conservation of energy, and total system angular momentum) into a set of algebraic expressions by estimating derivatives over the plasma scale length $a$. No profiles are assumed or integrated. Analytic equations are then used to complete all terms in this set of equations, with empirically-supported arguments for strong turbulence suppression by sheared $E\times B$ flow motivating the use of classical perpendicular losses.

\begin{figure}[H]
    \centering\
    \includegraphics[scale=.48]{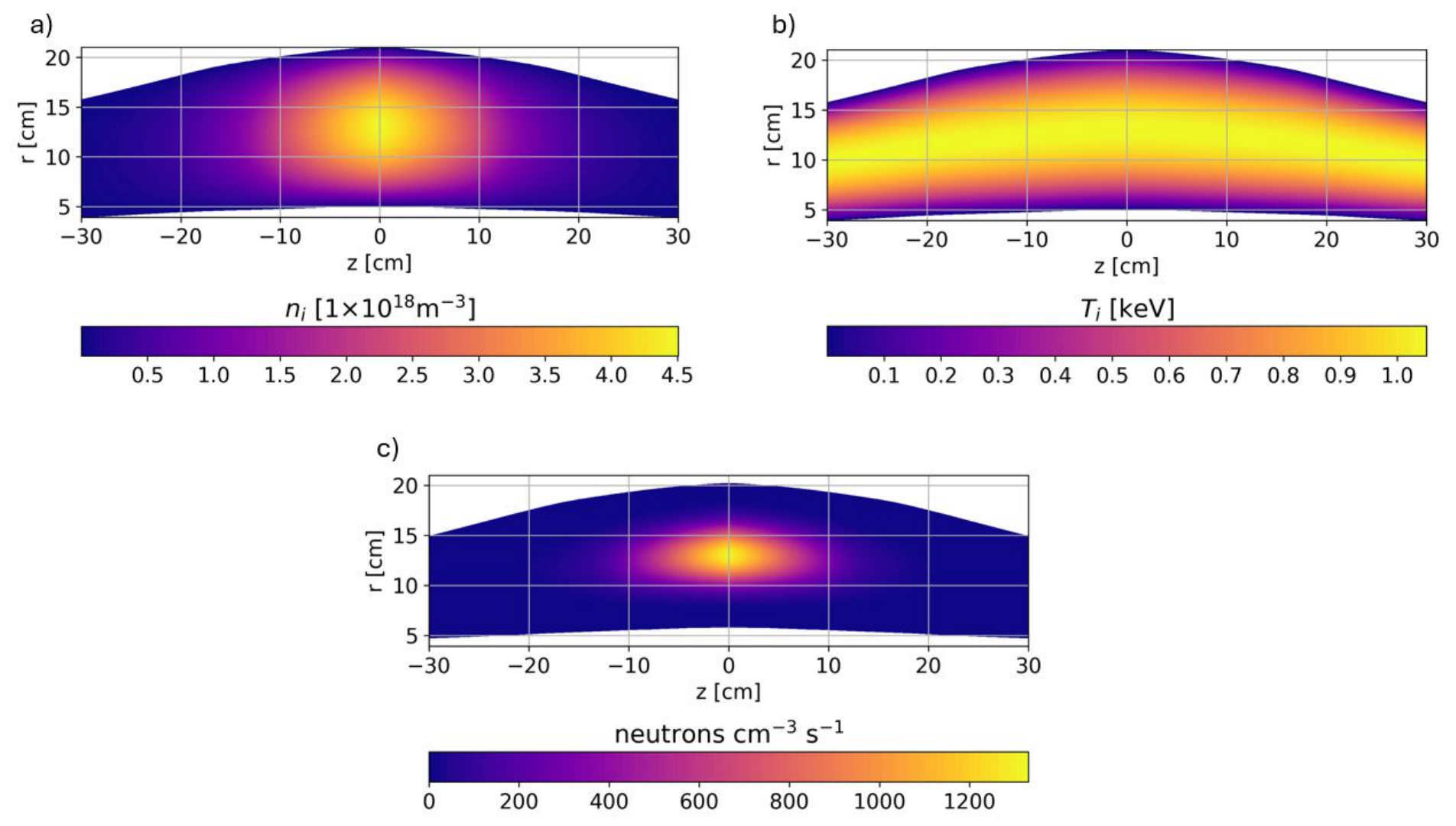}
    \caption{Example density and temperature profile mapping for neutron source calculation for Mach = 6.4. a) Ion density profile for $n_{i, \mathrm{avg}}$ = 3 $\times$ 10$^{18}$ m$^{-3}$ mapped axially in the combined centrifugal and Pastukhov potentials, and radially with a parabolic profile. b) Ion temperature with  $T_{i, \mathrm{avg}}$ = 0.7 keV is assumed to be a flux function that radially varies with a parabolic profile. c) The resulting neutron emissivity.}
    \label{fig:profiles}
\end{figure}

We did, however, assume profiles for estimating the neutron rate from the model. The ion density and temperature produced by MCTrans++ are treated as average values at the midplane, and are mapped along flux surfaces axially in the centrifugal plus Pastukhov potentials based on Equations 2.9 and 2.20 in \cite{mctrans}. Radially, a parabolic profile varying like

\begin{equation}
    1 - \frac{(r_{\mathrm{avg}} - r)^2}{ ((r_{\mathrm{outer}} - r_{\mathrm{inner}})/2)^2}
\end{equation}
\newline
 \noindent is used, which is less physically motivated than the axial transport, but agrees with MCTrans++'s assumption of radial profile gradients on the order of $a = (r_{\mathrm{outer}} - r_{\mathrm{inner}})/2$ from classical transport perpendicular to flux surfaces. These profiles are used to get local fusion D(d, n)$^3$He reactivity $R_{DD} = \frac{1}{2} n_D^2 \langle \sigma v \rangle_{DD}$, which is then integrated over the plasma volume to give a total DD neutron production rate. Example ion density and temperature profiles are shown in Figure \ref{fig:profiles} for Mach number $M$ = 6.4, $n_{i, \mathrm{avg}}$ = 3$\times$10$^{18}$ m$^{-3}$, $T_{i, \mathrm{avg}}$ = 0.7 keV. Fig.~\ref{fig:profiles}~a) shows the spatially varying $n_i$, and Fig.~\ref{fig:profiles} b) the spatially varying ion temperature, which allows for calculating spatially varying $\langle \sigma v \rangle_{DD}$ and the neutron emissivity profile shown in c).

\subsection{Plasma Operational Contours}

\begin{figure}[H]
    \centering\
    \includegraphics[scale=0.58]{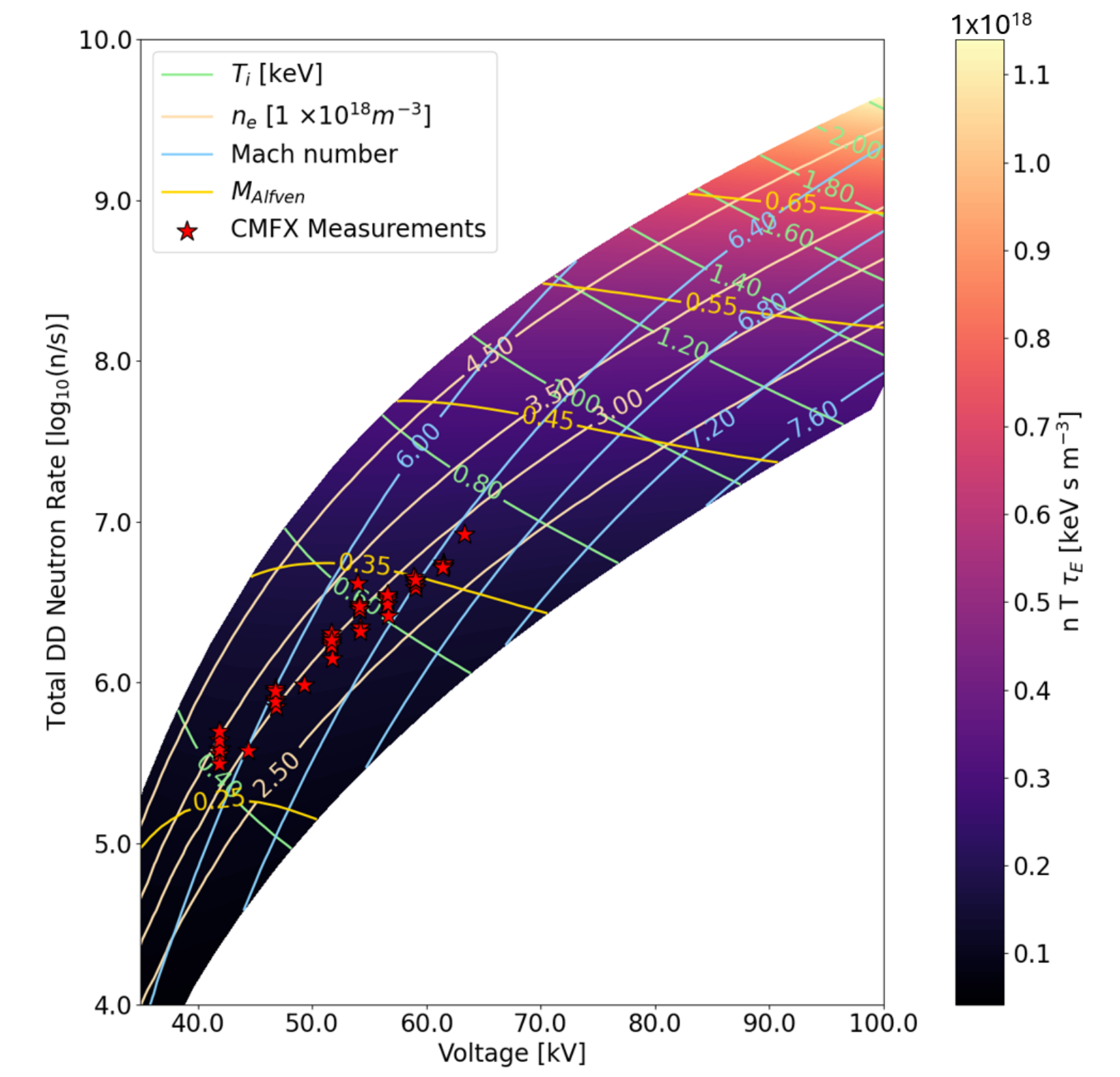}
    \caption{Contour plots produced by running MCTrans++ on 900 density-voltage combinations within the voltage capabilities of CMFX's 100 kV DC power supply. Shown are contours of various device parameters: the triple product $n_i T_i \tau_E$ predicted by the MCTrans++ model is shown as the color-fill behind the contours. The abscissa and ordinate are chosen to correspond to experimental observables: delivered voltage and total D-D fusion neutron rate. Overlaid with red stars are the experimentally measured discharges, with error bars excluded for clarity. Note the DD neutron rate scale is logarithmic, and a constant neutral density of 2.4$\times10^{14}$ m$^{-3}$ is assumed. The only experimentally determined values are the red points.}
    \label{fig:popcon}
\end{figure}

In steady-state mode, MCTrans++ takes as input the electron density and supplied voltage, along with device and plasma geometry, magnetic field, $Z_{\mathrm{eff}}$, and neutral density. This allows for the construction of Plasma Operation Contours (POPCONs)~\cite{popcons_1982}, with other device parameters such as heat fluxes, ion and electron temperatures, fusion neutron rate, viscous heating, thermal Mach number, and Alfv\'{e}n Mach numbers as solved parameters. To produce a POPCON for the CMFX device, 900 MCTrans++ runs were performed scanning 30 voltages from 35 to 100 kV (corresponding to the range of set voltages used in these experiments, plus an extension to 100 kV, the maximum that can be delivered by CMFX's DC power supply), and 30 densities from 2 to 5$\times10^{18}$ m$^{-3}$. 

The input parameters were a deuterium plasma length of 60 cm, wall radius of 36.1 cm, plasma minimum and maximum radii of 5 cm and 21 cm, respectively, and a $Z_{\mathrm{eff}}$ of 3.0. The throat field was 3.0 T and the central cell field was 0.34 T. Neutral density is an important parameter, affecting the power required to maintain rotation: charge exchange collisions are enhanced by rotation up until about 1 keV \cite{mctrans}. The neutral density is not expected to remain constant throughout this parameter space. However, for the POPCON studied here, a neutral density of 2.4$\times10^{14}$ m$^{-3}$ is used which corresponds to the middle of the range of neutral densities inferred for the shots performed. In the next subsection, a more detailed explanation is given for how the neutral density can be inferred by using experimental measurements to inform the MCTrans++ model. 

To better compare with experiment, the POPCON was transformed from $n_e$ - voltage space into DD fusion neutron rate-voltage space, with set voltage as the abscissa and total neutron rate as the ordinate, shown in Figure \ref{fig:popcon}. The neutron rate was found for a given MCTrans++ point via the profile mapping of the average ion density and temperatures, as discussed in the previous subsection and in Fig.~\ref{fig:profiles}. The background color fill is the triple product $n_D T_i \tau_E$. Plotted over the MCTrans++ model predictions are red points corresponding to the total neutron yield rate measurements described in Sections~\ref{sec:calib} and \ref{sec:performance}. It should be stressed that only the red points were measured, and all other contours in Figure \ref{fig:popcon} are modeled by MCTrans++ and the profile mapping in Fig. \ref{fig:profiles}. 

The POPCON suggests that should CMFX be upgraded to operate at the full 100 kV set voltage available from the DC power supply, 1.7 keV ion temperatures and triple products of nearly 1$\times10^{18}$ keV s m$^{-3}$ may be achievable when extrapolating along the 3$\times10^{18}$ m$^{-3}$ density contour where the data roughly lie. This would put CMFX performance near the EAST device and the GOL-3 mirror in the triple product versus temperature space \cite{Lawson_Review_Wurzel}. Next, MCTrans++ is used along with measurements to relax the constant neutral density assumption made in Fig.~\ref{fig:popcon} for a better estimate of the achieved ion density, temperature, and triple product for each experimental shot performed in this campaign.  

\subsection{Iterative Solve for Plasma Parameters}

An iterative method was developed to constrain the neutral density for consistency with measurements to be more confident in MCTrans++ predictions for shot performance. MCTrans++, when provided with a known magnetic field, an estimated plasma length of 0.6 m, and a conservative guess for $Z_{\mathrm{eff}}$ = 3, is left with three unknown independent variables: ion and electron temperature, electron density (assumed equal to ion density), and neutral density. By using the measured neutron rate averaged over the flattop from the 10-inch Scintillator, the measured average voltage, and the measured average power draw, the three unknowns can be constrained for each shot. 

\begin{figure}[H]
    \centering\
    \includegraphics[scale=0.36]{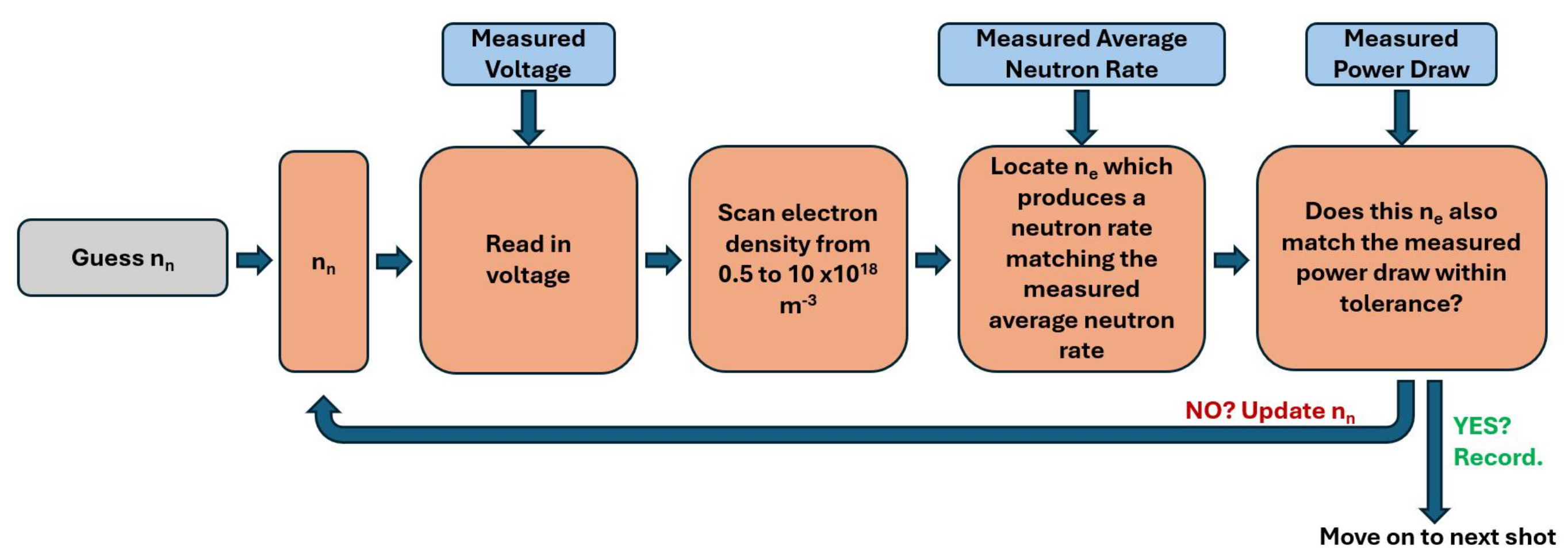}
    \caption{Flowchart showing the iterative method used to match MCTrans++ predictions to experimentally measured voltage, power draw, and average neutron rate. This scheme is performed for each shot.}
    \label{fig:flowchart}
\end{figure}

The method for achieving this is shown graphically in Figure \ref{fig:flowchart}. For each shot, the average flattop voltage, power draw, and neutron rate are recorded from measurements. Then, an initial guess neutral density is chosen; here, $1\times10^{14}$ m$^{-3}$ is used. This neutral density is used to scan MCTrans++, holding all variables constant except for the electron density, which is scanned from 0.5 to 10$\times10^{18}$ m$^{-3}$ at the shot's measured supply voltage. Next, the electron density which results in a neutron rate that matches the measured neutron rate is recorded, and its MCTrans++ predicted power draw is extracted. The predicted power draw is compared to the measured average flattop power draw. If the power draw predicted by MCTrans++ does not agree with the measured power draw, the neutral density guess is updated for the next scan using a Newton iteration method. This iteration continues until the power draw percent error between predicted and measured power draw was below a threshold value, chosen here to be 1$\times10^{-3}$. This scheme was repeated for every shot.

\begin{figure}[H]
    \centering\
    \includegraphics[scale=0.40]{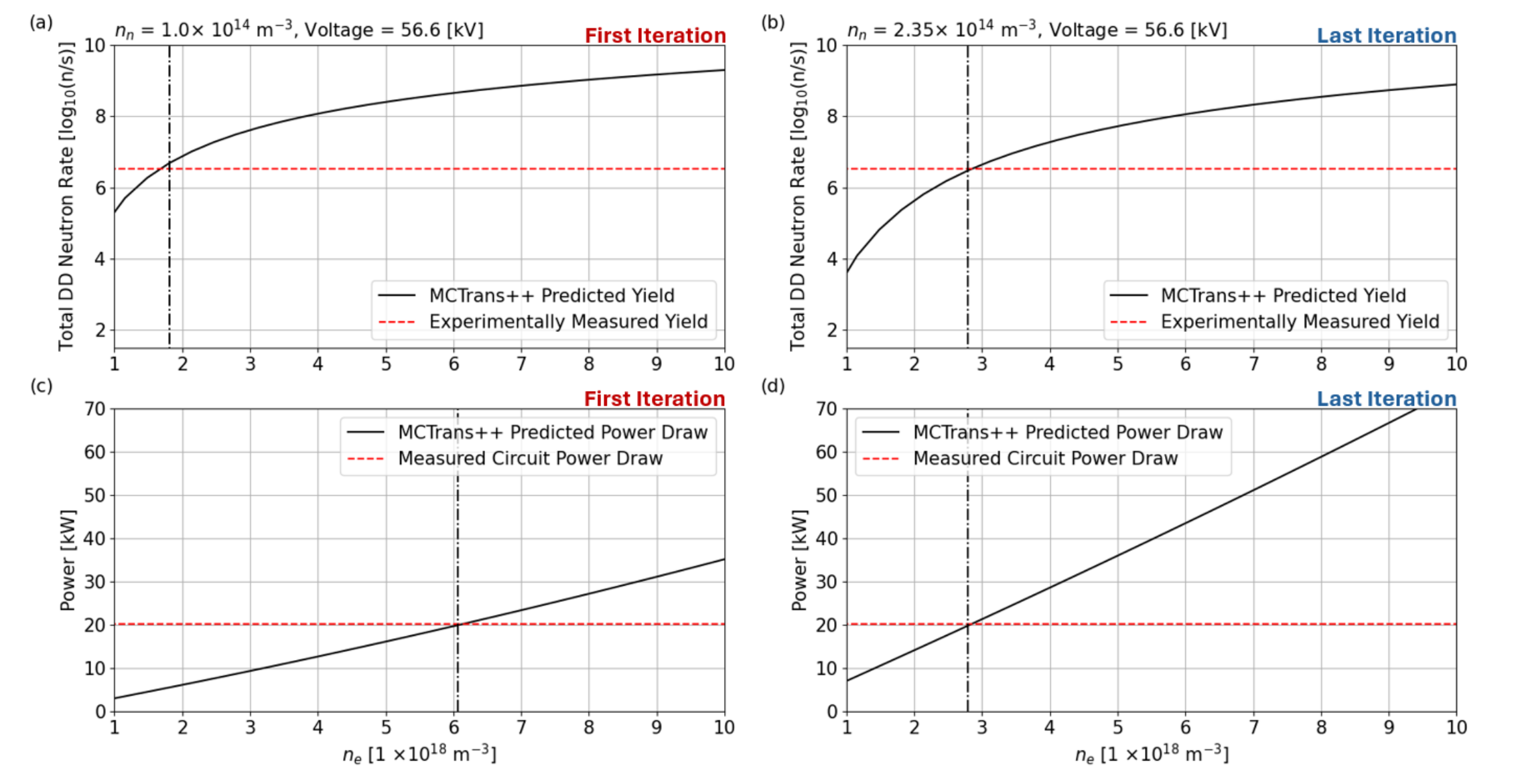}
    \caption{An example of the initial and final scan in the iteration scheme for a single shot. a) and c) show in solid black the neutron rate and power draw, respectively, predicted by MCTrans++ for this shot's initial neutral density guess of 1$\times10^{14}$ m$^{-3}$. In red dashed are the measured averaged neutron rate and the measured circuit power draw. Notably, the vertical dashed lines showing the intersection do not line up between a) and c), indicating the neutral density is not self-consistent. In b) and d), the same information is shown for the iteration's final converged neutral density of 2.35$\times10^{14}$ m$^{-3}$. The vertical black lines now align, showing convergence.}
    \label{fig:iteration}
\end{figure}

As an example of this method, shot 2071's first and last neutral density iteration are shown in Figure \ref{fig:iteration}. In Figure \ref{fig:iteration} a) and c), the neutron rate prediction and power draw prediction are shown in solid black as a function of electron density from a scan of 30 electron densities for the initial guess neutral density of 1$\times10^{14}$ m$^{-3}$. The measured neutron rate and measured power draw for this shot are shown with red dashed lines. Notably, the density where the intersection between the measured and predicted neutron rates and power draw occurs, marked with vertical black dashed lines, does not line up between a) and c), indicating the neutral density used is not consistent with measurements for this shot's voltage. After 12 iterations, the final neutral density for this shot converged to 2.35$\times10^{14}$ m$^{-3}$; the corresponding neutron rate and power draw curves are shown in Figure \ref{fig:iteration} b) and d). We see the dashed vertical black line where the predicted and measured dashed red line intersect line up between b) and d), demonstrating the convergence at the electron density of the intersection. 

This method was performed on all of the shots studied here, from which MCTrans++ predictions for other physical parameters, such as triple product, electron and ion temperatures, etc., shown in the next section, are made.

\subsection{Inferred Ion Temperature, Density, and Triple Product}
After running the iterative method discussed in the previous subsection for all shots, the resultant MCTrans++ inferences for various plasma properties are extracted. It is important to recognize the limitations of the accuracy of this process. The MCTrans++ model is 0D, and $Z_{\mathrm{eff}}$ and plasma length were chosen with values that are conservative but not measured or calculated, and radial profiles were assumed to be parabolic. 

\begin{figure}[H]
    \centering\
    \includegraphics[scale=0.55]{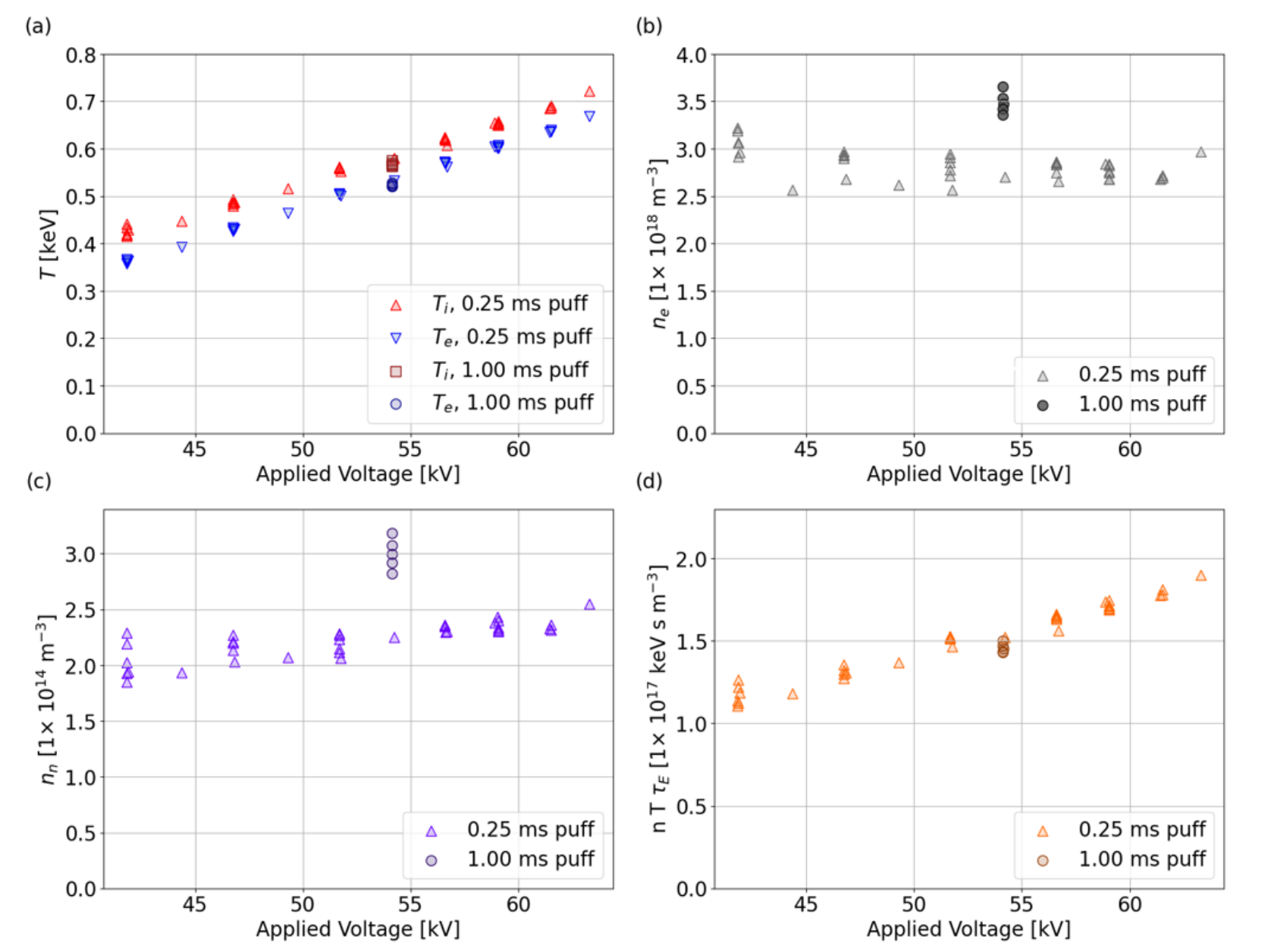}
    \caption{MCTrans++ model predictions for a)~ion and electron temperature, b)~electron density, c)~neutral density, and d)~fusion triple product. Values for both a 0.25 ms deuterium fueling puff and a 1.00 ms puff for a single set voltage are shown.}
    \label{fig:trends}
\end{figure}

The inferred temperatures in Figure~ \ref{fig:trends}~a) indicate that an average ion temperature of 720 eV and average electron temperature of 670 eV were achieved for the highest voltage shot undertaken in this campaign. Ion temperatures remained above electron temperatures for all experiments, as expected for low temperature centrifugal mirror operation \cite{mctrans}. In Fig.~\ref{fig:trends} b), the electron density is shown, which displays a relatively constant trend with voltage. Roughly 2.8$\times$10$^{18}$ m$^{-3}$ is inferred. The neutral density in Figure \ref{fig:trends} c) increases slightly with voltage, with average values around 2.4$\times$10$^{14}$ m$^{-3}$ (this was the neutral density used in making the POPCON in Figure \ref{fig:popcon}). Lastly, Fig.~\ref{fig:trends} d) shows an increasing triple product with voltage, reaching 1.9$\times$10$^{17}$ keV s m$^{-3}$ for the most performant discharge.  

In the shots performed, the gas puff length was kept constant except for shots with set voltage of 55 kV (the average delivered voltage being slightly less than 55 kV, around 54 kV). For these shots, the puff length was 1 ms, four times longer than the other shots in this data set. These points are marked in Figure \ref{fig:trends}. Noticeably, the longer puff length had a smaller effect on the temperature trend in Figure \ref{fig:trends} a) and the triple product trend in d). Given that the electron density in b) is increased for the longer puff, the confinement time must have been reduced for the longer puff shots. This confinement time reduction is likely due to increased charge exchange losses. 

\begin{figure}[H]
    \centering\
    \includegraphics[scale=0.4]{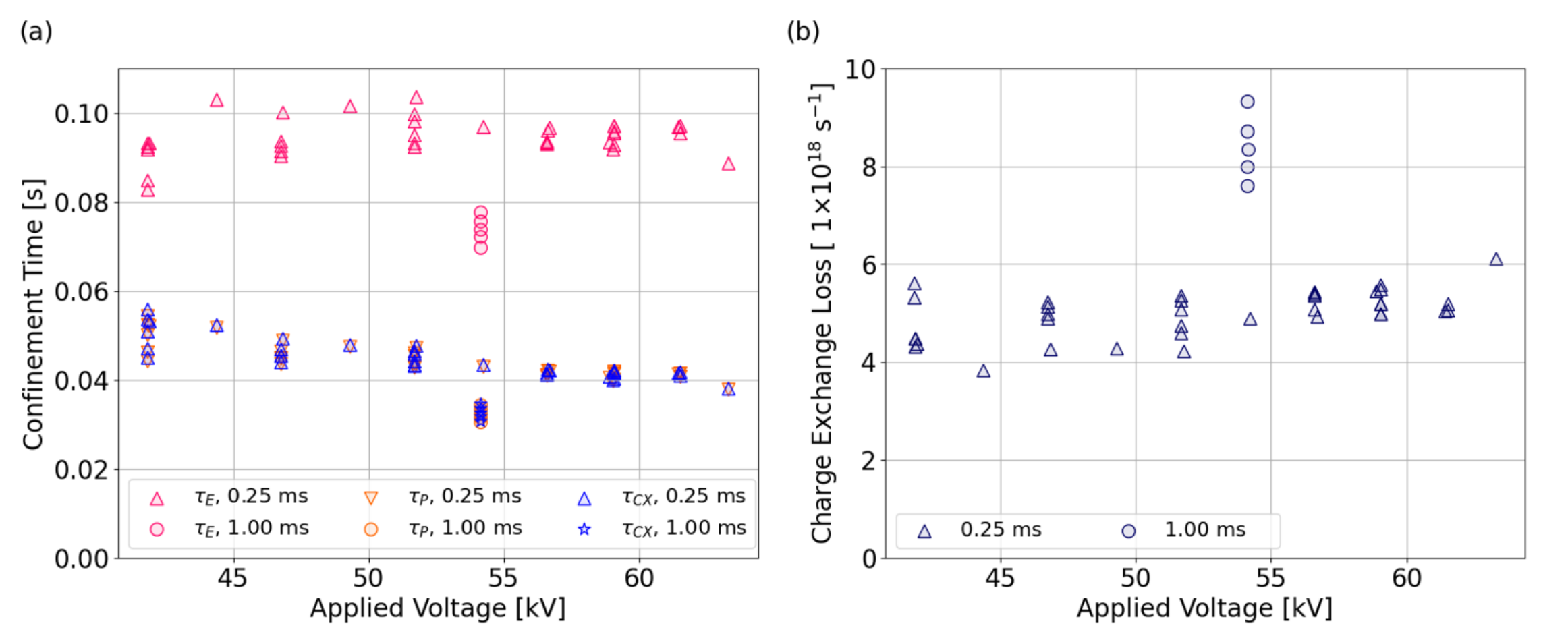}
    \caption{a) MCTrans++ predicted energy confinement time $\tau_{E}$ (pink), particle confinement time $\tau_{P}$ (orange), and charge exchange confinement time $\tau_{CX}$ (blue) for all shots across the 0.25 and 1.00 ms gas puffs. b) Total device charge exchange loss rate for all shots. 1 ms longer-puff shots are marked with the circles or stars symbol, the rest are 0.25 ms shots, marked with triangles. Enhanced charge exchange losses for the longer-puff shots explain the trend for the longer puff shots in Figure \ref{fig:trends} b), c), d)}
    \label{fig:confinement_time}
\end{figure}

Plots of the energy confinement time, particle confinement time, and charge exchange confinement time inferred by the model for the shots performed here are shown in Fig.~\ref{fig:confinement_time} a). The particle confinement time (orange) nearly exactly tracks the charge exchange confinement time (blue), with the longer 1 ms puff shots marked with circles and stars, demonstrating a clear reduction. The reduced energy confinement time (pink) tracks the reduction in the charge exchange confinement time, indicating the reduction is due to energy carried away in the fast neutral product particles of the charge exchange reactions. This dip confirms the explanation that lower energy confinement time due to enhanced charge exchange loss in the longer-puff shots is responsible for the small change in triple product between the 0.25 and 1.00 ms shots. Fig.~\ref{fig:confinement_time} b) shows the overall charge exchange rate, with an obvious enhancement for the longer-puff shots, marked with circles.  


\section{Conclusion}
\label{sec:conclusion}

In this work, we used two different absolutely calibrated neutron detection techniques to measure the total neutron yield rate of the Centrifugal Mirror Fusion Experiment for the first time. The first of these techniques involved the use of three liquid scintillator detectors: two well characterized 2-inch detectors and one large 10-inch detector. Two computational models, one of neutron transport from the plasma to the detectors (OpenMC) and one of the detector response (Geant4 + empirical data), were used to perform an \textit{in silico} absolute calibration of the 2-inch detectors. These detectors were then used to cross-calibrate the much larger 10-inch detector (see Fig.~\ref{fig:10inchCrossCalib}), which was used to study low-yield discharges and temporal dynamics (see Figs.~\ref{fig:60kV shots},~\ref{fig:fueling-study}). The results of these yield measurements were then compared with those made by an \textit{in situ} calibrated $^3$He tube, where a $^{252}$Cf source was positioned at six locations along an arc to model the neutron emission from CMFX. Excellent linearity was observed between the $^3$He and liquid scintillator measurements. A linear fit found no substantial systematic bias between the two measurement approaches, with a slope of 0.98 and an intercept $<2\%$ of the maximum yield rate measured. Fig.~\ref{fig:Scint-3He comparison} shows this linear fit against the $^3$He and scintillator measured yields for 42 discharges on CMFX.

Some sources of uncertainty in the scintillator measurement, like model bias, were neglected due to the large Poisson uncertainty dominating the 2-inch absolute yield measurements. This choice was also motivated by the neutron transport from source to detector being mostly direct (see Fig.~\ref{fig:openmcSpectrum}), and by the strong agreement between synthetic and experimental pulse height spectra (see Fig.~\ref{fig:2in_cumulative_phs}). This choice was later supported by the excellent agreement found between the \textit{in silico} scintillator calibration and \textit{in situ} $^3$He calibration discussed above, indicating no large source of systematic uncertainty was ignored. Future work should look to better quantify these sources of model uncertainty so that this method may be applied independent of \textit{in situ} comparisons. Nonetheless, it is clear that accurate total neutron yield measurements can be made with liquid scintillators if neutron transport and detector response are adequately modeled.

These absolutely calibrated detectors were then used to study the fusion neutron emission of CMFX. It was found that CMFX has high shot repeatability, including for neutron yield rate (see Fig.~\ref{fig:60kV shots}). A fueling study was also conducted, and three different fueling schemes compared. It was found that spreading fueling out into two smaller gas puffs instead of one larger puff provided improved performance, motivating further investigations of fueling optimization (see Fig.~\ref{fig:fueling-study}). The data from 28 shots at a range of voltages was then compiled into a voltage versus yield rate curve, to which an exponential function was fit (see Fig. \ref{fig:voltage_yield}). The maximum average neutron yield rate observed was 8.4$\times 10^{6}$ $\pm$ 7.0$\times 10^{5}$ neutrons per second at a voltage of 65 kV.

To further our understanding of these results, the 0D centrifugal mirror code MCTrans++ was used to estimate other parameters of interest like ion temperature and density. This modeling suggests that in its highest performing discharges, CMFX has an average ion temperature of 720 eV and electron density of $2.8 \times 10^{18}$ m$^{-3}$, with a peak triple product of $1.9 \times 10^{17}$ keV s m$^{-3}$. This places CMFX in a similar performance regime to the Madison Symmetric Torus, the C-2W field reverse configuration, and the ST tokamak \cite{Lawson_Review_Wurzel}. This result is achieved at a set voltage of 65 kV, although the CMFX power supply is capable of 100 kV. MCTrans++ modeling indicates that CMFX could achieve triple products as high as $1 \times 10^{18}$ keV s m$^{-3}$ at 100 kV, strongly motivating the upgrade of the machine to be capable of stable operation at this voltage. Further extrapolations of rotating mirror performance to DT reactor regimes have been performed using the MCTrans++ code, and can be found in Ref.~\cite{mctrans}. These results strongly indicate that the centrifugal mirror is worthy of further study as a possible fusion reactor configuration.

\section*{Acknowledgments}

CMFX is supported by ARPA-E Grant No. DE-AR0001270. \\

\noindent 2-inch liquid scintillator detectors, CAEN DT5730SB digitizer, and CAEN DT1471ET HV power supply used in this work acquired by MIT PSFC under Commonwealth Fusion Systems RPP 031. \\

\noindent Thanks to Ian Abel for assistance in the use of MCTrans++ for this work. \\

\noindent Thanks to Zach Hartwig for providing access to the large 10-inch scinillator used in this work.

\bibliographystyle{ieeetr}
\bibliography{references.bib}

\end{document}